\documentstyle[axodraw,psfig,preprint,prd,aps]{revtex}
\input epsf.sty

\begin{document}
\hbadness=10000 \pagenumbering{arabic}

\preprint{{\vbox{\hbox{ HEP-PH/0107165,
IPAS-HEP-01-k003, NCKU-HEP-01-06}}}} \vspace{1.5cm}

\title{\Large \bf
Perturbative QCD analysis of $B \rightarrow \phi K$ decays
and power counting}
\date{
\today%
}

\vskip2.0cm
\author{\large \bf Chuan-Hung Chen$^{a}$\footnote{Email: 
chchen@phys.nthu.edu.tw}, 
Yong-Yeon Keum$^{b}$\footnote{Email: keum@phys.sinica.edu.tw} 
and Hsiang-nan Li$^{a,c}$\footnote{Email: hnli@mail.ncku.edu.tw}}

\vskip2.0cm

\address{ $^{a}$ Department of Physics, National Cheng-Kung University,\\
Tainan, Taiwan 701, Republic of China}
\address{ $^{b}$ Institute of Physics, Academia Sinica,
Taipei, Taiwan 115, Republic of China}
\address{ $^{c}$ Physics Division, National Center for Theoretical
Sciences,\\
Hsinchu, Taiwan 300, Republic of China}

\maketitle

\newpage

\begin{abstract}
We investigate exclusive nonleptonic $B$ meson decays $B\to\phi K$ in 
perturbative QCD formalism. It is shown that the end-point (logarithmic
and linear) singularities in decay amplitudes do not exist, after $k_T$
and threshold resummations are included. Power counting for
emission and annihilation topologies of diagrams, including both
factorizable and nonfactorizable ones, is discussed with
Sudakov effects taken into account. Our predictions for the branching
ratios $B(B\to\phi K)\sim 10 \times 10^{-6}$ are larger than
those ($\sim 4 \times 10^{-6}$) from the factorization approach because
of dynamical enhancement of penguin contributions. Whether this
enhancement is essential for penguin-dominated modes can be justified
by experimental data.
\end{abstract}

\vskip2.0cm
{\bf PACS number(s): 12.38.Bx, 13.35.Hw, 12.38.Qk, 11.10.Hi}

\newpage

\section{INTRODUCTION}

Perturbative QCD (PQCD) factorization theorem for the semileptonic decay
$B\to\pi l\bar\nu$ has been proved in \cite{L4}, which states that the
soft divergences in the $B\to\pi$ form factor can be factorized into a
light-cone $B$ meson distribution amplitude and the collinear divergences
can be factorized into a pion distribution amplitude order by order. 
The remaining finite contribution is assigned into a hard amplitude,
which is calculable in perturbation theory. A meson distribution
amplitude, though not calculable, is universal, since it absorbs
long-distance dynamics, which is insensitive to specific decays of the $b$
quark into light quarks with large energy release. The universality of
nonperturbative distribution amplitudes is one of the important
ingredients of PQCD factorization theorem. Because of this universality,
one extracts distribution amplitudes from experimental data, and then
employ them to make model-independent predictions for other processes. In
this paper we shall assume that PQCD factorization theorem holds for
two-body nonleptonic $B$ meson decays, to which there is no difficulty
to generalize the proof in \cite{L4}. The one-loop proof for the
PQCD factorization of two-body decays has been given in \cite{CL}.

The PQCD formalism for the charmed decays $B\to D^{(*)}\pi(\rho)$ 
\cite{CL,YL} is restricted to twist-2 (leading-twist) distribution 
amplitudes. For charmless decays such as $B\to K\pi$, $\pi\pi$ and
$KK$ \cite{{Keum:2001ph},{Keum:2001wi},{Keum:2001ms},LUY,CHL}, 
contributions from two-parton twist-3 (next-to-leading-twist) 
distribution amplitudes are introduced via the penguin operators $O_{5-8}$ 
in the effective Hamiltonian for weak decays. It has been argued \cite{L6} 
that two-parton twist-3 contributions are in fact not suppressed by 
a power of $1/M_B$, $M_B$ being the $B$ meson mass. A compelte 
leading-power PQCD analysis of the heavy-to-light $B\to\pi$, $\rho$ form 
factors, including both twist-2 and twist-3 contributions, has been
performed in \cite{TLS}. There exist many other higher-twist sources in
$B$ meson decays, whose contributions are indeed down by a power of
$1/M_B$. These sources include the $B$ meson and $b$ quark mass difference 
${\bar\Lambda}=M_B-m_b$, the light quark masses $m_u$, $m_d$ and $m_s$, 
and the light pseudo-scalar meson masses $M_\pi$ and $M_K$. Those from 
three-parton distribution amplitudes are further suppressed by the
coupling constant $\alpha_s$. All these sub-leading contributions will 
be neglected in the current formalism. 

In this work we shall perform a PQCD analysis of the $B\to\phi K$ decays
up to corrections of $O(\bar\Lambda/M_B)$. These modes involve different 
topologies of diagrams, such as factorizable (also nonfactorizable) 
emission and annihilaton. We predict the branching ratios and CP 
asymmetries of the modes 
\begin{eqnarray}
B^0_d\to \phi K^0\;,\;\;\;B^\pm\to \phi K^\pm\;.
\end{eqnarray}
It will be found that two-parton twist-3 contributions are comparable to 
twist-2 ones as expected. Our predictions for the braching ratios
\begin{eqnarray}
B(B^\pm\to\phi K^\pm) &=& (10.2^{+3.9}_{-2.1})  \times 10^{-6}\; 
\nonumber \\
B(B_d^0\to\phi K^0) &=& (9.6^{+3.7}_{-2.0}) \times 10^{-6}\;,
\label{phikb}
\end{eqnarray}
are larger than those from the factorization approach \cite{BSW} and from
the QCD factorization approach \cite{BBNS}, which are located within
$4.3^{+3.0}_{-1.4}\times 10^{-6}$ \cite{HMW,CY} with the uncertainty 
arising mainly from the inclusion of annihilation contributions \cite{CY}.
The mechanism responsible for the larger branching ratios in the PQCD
formalism is dynamical enhancement of penguin contributions
\cite{{Keum:2001ph},{Keum:2001wi}}. Note that the current experimental
data of $B(B^\pm\to \phi K^\pm)$, 
\begin{eqnarray}
{\mathrm CLEO \cite{CLEO}}: & &\hspace{20mm}
 (5.5^{+2.1}_{-1.8}\pm 0.6)\times 10^{-6},
\nonumber\\
{\mathrm Belle \cite{Belle-1}}: & & \hspace{20mm}
(10.6^{+2.1}_{-1.9} \pm 2.2)\times 10^{-6}\;,
\nonumber \\
{\mathrm BaBar \cite{Babar}}: & & \hspace{20mm}
(7.7^{+1.6}_{-1.4}\pm 0.8)\times 10^{-6}\;,
\end{eqnarray}
and those of $B(B^0 \to \phi K^0)$, 
\begin{eqnarray}
{\mathrm CLEO \cite{CLEO}}: & & \hspace{20mm} < 12.3 \times 10^{-6},
\nonumber\\
{\mathrm Belle \cite{Belle-1}}: & & \hspace{20mm} 
(8.7^{+3.8}_{-3.0}\pm 1.5) \times 10^{-6}\;,
\nonumber \\
{\mathrm BaBar \cite{Babar}}: & & \hspace{20mm}
(8.1^{+3.1}_{-2.5}\pm 0.8)\times 10^{-6}\;,
\end{eqnarray}
are still not very consistent with each other. 

The PQCD factorization formulas for $B\to\phi K$ decay amplitudes
have been derived independently in \cite{Mi}. Here we shall further 
discuss the power behavior of the factorizable emission and 
annihilation amplitudes, and the nonfactorizable amplitudes in $1/M_B$, 
whose relative importance is given by
\begin{eqnarray}
{\rm emission} : {\rm annihilation} : {\rm nonfactorizable} 
=1 : \frac{2m_0}{M_B} : \frac{\bar\Lambda}{M_B}\;,
\label{rim}
\end{eqnarray}
with $m_0$ being the chiral symmetry breaking scale.
In the heavy quark limit the annihilation and nonfactorizable amplitudes
are indeed power-suppressed compared to the factorizable emission ones.
Therefore, the PQCD formalism for two-body charmless nonleptonic $B$
meson decays coincides with the factorization approach as
$M_B\to\infty$. We shall also explain why the annihilation amplitudes
are mainly imaginary and investigate theoretical uncertainty
in the PQCD approach.

Dynamical enhancement is a unique feature of the PQCD approach, which 
does not exist in the factorization or QCD factorization approach. We
argue that the $B\to\phi K$ modes are more appropriate for testing this
mechanism in penguin-dominated nonleptonic $B$ meson decays compared 
to the $B\to K\pi$ decays \cite{L6}. The large $B\to K\pi$ branching 
ratios may not be regarded as an evidence of dynamical enhancement: 
they can also be achieved by chiral enhancement (the kaon is a 
pseudo-scalar meson) and by choosing a large unitarity angle 
$\phi_3 \sim 120^o$ \cite{WS}, 
which leads to constructive interference between penguin and emission 
contributions. The $B\to \phi K$ modes are not chirally enhanced, 
because $\phi$ is a vector meson, and insensitive to the variation of 
the angle $\phi_3$, because they are pure penguin processes. If the data 
of $B(B\to \phi K)$ are settled down at values around $10\times 10^{-6}$ 
in the future, dynamical enhancement will gain a strong support.

On the other hand, precise measurement of the CP asymmetry in the 
$B \to \phi K$ decays is important for new physics search and 
for the determination of the unitarity angle $\phi_1$ with high degree 
of accuracy \cite{LP,GIW}. This measurement is experimentally accessible 
at the early stage of the asymmetric $B$ factories. The $B\to \phi K$
decays arise from penguin (loop) effects, while the $B\to J/\psi K$ decays
involve only tree amplitudes. The search for different CP asymmetries in
the $B \to J/\psi K_s$ and $\phi K_s$ decays, with the common source from
$B^0$-$\bar{B}^0$ mixing, provides a promising way to discover
new physics \cite{GW,NEW}: a difference of
$|A_{CP}(J/\psi K_s) - A_{CP}(\phi K_s)| > 5 \%$ would be an indication
of new physics. This subject will be addressed elsewhere \cite{BK}.
Furthermore, the $\phi$ and $J/\psi$ mesons are bound states of the
$s{\bar s}$ and $c{\bar c}$ quarks. It is also interesting to compare
their branching ratios, which reflect the mass effect from charm quarks.

We demonstrate the importance of $k_T$ and threshold resummations by 
studying the $B\to \pi, K$ transition form factors in Sec.~II. The power 
counting and the factorization formulas for various topologies of
amplitudes are given in Sec.~III. The numerical analysis is performed in
Sec.~IV. Section V is the conclusion. Twist-2 and two-parton twist-3
distribution amplitudes for the kaon and for the $\phi$ meson are defined
in the Appendix.

\vskip1.0cm

\section{SUDAKOV SUPPRESSION}

In this section we briefly review the importance of $k_T$ and threshold  
resummations for an infrared finite PQCD calculation of heavy-to-light
transition form factors \cite{TLS}. Consider the leading diagram shown
in Fig.~1 for the $B\to\phi K$ decays in the kinematic region with a
fast-recoil kaon. The $B$ meson momentum $P_1$, the $\phi$ meson
momentum $P_2$ and the longitudinal polarization vector $\epsilon$, and
the kaon momentum $P_3$ are chosen, in light-cone coordinates, as
\begin{eqnarray}
P_1=\frac{M_B}{\sqrt{2}}(1,1,{\bf 0}_T)\;,\;\;\;
P_2&=&\frac{M_B}{\sqrt{2}}(1,r_\phi^2,{\bf 0}_T)\;,\;\;\;
P_3=\frac{M_B}{\sqrt{2}}(0,1-r_\phi^2,{\bf 0}_T)\;,\;\;\;
\nonumber \\
\epsilon &=&\frac{1}{\sqrt{2}r_\phi}(1,-r_\phi^2,{\bf 0}_T)\;,
\label{pa}
\end{eqnarray}
with the ratio $r_\phi=M_{\phi}/M_B$, $M_\phi$ being the $\phi$ meson
mass. At the end of the derivation of the factorization formulas,
the terms $r_\phi^2\sim 0.04$ in the above kinematic variables will be
neglected. We treat the kaon as a massless particle, and define the ratio
$r_K=m_0/M_B$ for the kaon, which will appear in the normalization of the
twist-3 kaon distribution amplitudes. The $B$ meson is at rest under the
above parametrization of momenta.

\begin{center} \hspace{10mm}
\vspace{50pt} \hfill \\
\begin{picture}(90,0)(90,25)
\GOval(30,-10)(20,10)(0){0.5}
\GOval(170,-10)(20,10)(0){0.5}
\GOval(100,-10)(20,10)(0){0.5}
\GOval(100,40)(5,20)(0){0.5}
\ArrowLine(100,10)(30,10)
\ArrowLine(170,10)(100,10)
\ArrowLine(80,40)(100,10)
\ArrowLine(100,10)(120,40)
\ArrowLine(30,-30)(100,-30)
\ArrowLine(100,-30)(170,-30)
\GBoxc(100,10)(7,7){0}
\Text(73,15)[]{$b$}
\Text(50,-37)[]{$k_1$}
\Text(80,30)[]{$s$}
\Text(120,20)[]{$k_2$}
\Text(120,30)[]{$s$}
\Text(135,15)[]{$s$}
\Text(100,-37)[]{$d,u$}
\Text(150,-37)[]{$k_3$}
\ArrowLine(-10,-10)(30,-10)
\ArrowLine(170,-10)(210,-10)
\ArrowLine(100,40)(100,70)
\Text(-25,-10)[]{$B(P_1)$}
\Text(100,-10)[]{$H$}
\Text(225,-10)[]{$K(P_3)$}
\Text(100,85)[]{$\phi(P_2,\epsilon)$}
\end{picture} 
\end{center}

\vskip2cm
\begin{center}
{\bf FIG. 1}:
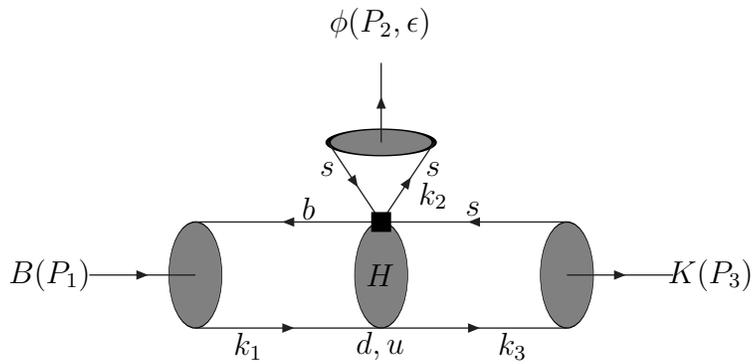
{ Leading contribution to the $B \to \phi K $ decays, where
$H$ denotes the hard amplitude and $k_i$, $i=1$, 2, and 3,
are the parton momenta.}
\end{center}

\subsection{$k_T$ and threshold resummations}

It has been known that the lowest-order diagram in Fig.~2(a) for  
the $B\to\pi$ form factor $F^{B\pi}$ is proportional to $1/(x_1 x_3^2)$
without including parton transverse momenta $k_T$, where
$x_1=k_1^+/P_1^+$ ($x_3=k_3^-/P_3^-$) is the momentum fraction
associated with the spectator quark on the $B$ meson (pion) side. If the
pion distribution amplitude vanishes like $x_3$ as $x_3\to 0$ (in the
twist-2 case), $F^{B\pi}$ is logarithmically divergent. If the pion
distribution amplitude is a constant as $x_3\to 0$ (in the twist-3 case),
$F^{B\pi}$ even becomes linearly divergent. These end-point singularities
have caused critiques on the perturbative evaluation of the $B\to\pi$
form factor. Several methods have been proposed to regulate the above
singularities. An on-shell $b$ quark propagator has been subtracted from
the hard amplitude as $x_3\to 0$ in \cite{ASY}. However, this subtraction
renders the lepton energy spectrum of the semileptonic decay
$B\to\pi l\bar\nu$ vanishes as the lepton energy is equal to half of its
maximal value. Obviously, this vanishing is unphysical, indicating that
the subtraction may not be an appropriate way to regulate the singularity.
The subtraction also leads to a value of $F^{B\pi}$ at maximal recoil,
which is much smaller than the expected one 0.3. A lower bound of $x_3$
of $O(\bar\Lambda/M_B)$ has been introduced in \cite{SHB} to make the
convolution integral finite. However, the outcomes depend on the cutoff
sensitively, and PQCD loses its predictive power.

\begin{center}
\vspace{50pt} \hfill \\
\begin{picture}(90,0)(90,25)
\GOval(30,-10)(20,10)(0){0.5}
\GOval(150,-10)(20,10)(0){0.5}
\ArrowLine(100,10)(30,10)
\ArrowLine(150,10)(100,10)
\ArrowLine(30,-30)(100,-30)
\ArrowLine(100,-30)(150,-30)
\Gluon(60,10)(60,-30){5}{5}
\GBoxc(100,10)(7,7){0}
\ZigZag(100,10)(100,40){5}{5}
\Text(50,15)[]{$b$}
\Text(50,-37)[]{$k_1$}
\Text(120,15)[]{$d,s$}
\Text(120,-37)[]{$k_3$}
\ArrowLine(-10,-10)(30,-10)
\ArrowLine(150,-10)(180,-10)
\Text(-25,-10)[]{$B$}
\Text(200,-10)[]{$\pi,K$}
\Text(100,-60)[]{$(a)$}
\end{picture} \hspace{40mm}
\begin{picture}
(90,0)(90,25)
\GOval(60,-10)(20,10)(0){0.5}
\GOval(170,-10)(20,10)(0){0.5}
\ArrowLine(100,10)(60,10)
\ArrowLine(170,10)(100,10)
\ArrowLine(60,-30)(100,-30)
\ArrowLine(100,-30)(170,-30)
\Gluon(130,10)(130,-30){5}{5}
\GBoxc(100,10)(7,7){0}
\ZigZag(100,10)(100,40){5}{5}
\Text(85,15)[]{$b$}
\Text(85,-37)[]{$k_1$}
\Text(150,15)[]{$d,s$}
\Text(150,-37)[]{$k_3$}
\ArrowLine(30,-10)(60,-10)
\ArrowLine(170,-10)(210,-10)
\Text(20,-10)[]{$B$}
\Text(225,-10)[]{$\pi,K$}
\Text(100,-60)[]{$(b)$}
\end{picture} 
\end{center}

\vskip3cm
\begin{center}
{\bf FIG. 2}:
{Lowest-order diagrams for the $B \to \pi, K$ transition form factors.}
\end{center}

A self-consistent prescription has been proposed in \cite{LY1}, where
parton transverse moemta $k_T$ are retained in internal particle
propagators involved in a hard amplitude. In the end-point region the
invariant mass of the exchanged gluon is only $O(\bar\Lambda^2)$ without
including $k_T$. The inclusion of $k_T$ brings in large double logarithms 
$\alpha_s\ln^2(k_T/M_B)$ through radiative corrections, which should be
resummed in order to improve perturbative expansion. $k_T$ resummation
\cite{CS,BS} then gives a distribution of $k_T$ with the average
$\langle k_T^2\rangle\sim O(\bar\Lambda M_B)$ for $M_B\sim 5$ GeV. The
off-shellness of internal particles then remains of $O(\bar\Lambda M_B)$
even at the end point, and the singularities are removed. Hence, it can
not be self-consistent to treat $k_T$ as a higher-twist effect as the
end-point region is important. The expansion parameter
$\alpha_s(\bar\Lambda M_B)/\pi\sim 0.13$ is also small enough to justify
PQCD evaluation of heavy-to-light form factors
\cite{{Keum:2001ph},{Keum:2001wi}}. This result is so-called Sudakov 
suppression on the end-point singularities in exclusive processes
\cite{LS}.

The above discussion applies to the $B\to K$ form factor and to the
$B\to \phi K$ decays. $k_T$ resummation of large
logarithmic corrections to the $B$, $\phi$ and $K$ meson distribution 
amplitudes lead to the exponentials $S_{B}$, $S_{\phi}$ and $S_{K}$, 
respectively:
\begin{eqnarray}
S_{B}(t)&=&\exp\left[-s(x_{1}P_{1}^{+},b_{1})
-2\int_{1/b_{1}}^{t}\frac{d{\bar{\mu}}} {\bar{\mu}}
\gamma (\alpha _{s}({\bar{\mu}}^2))\right]\;,
\nonumber \\
S_{\phi }(t)&=&\exp\left[-s(x_{2}P_{2}^{+},b_{2})
-s((1-x_{2})P_{2}^{+},b_{2})
-2\int_{1/b_{2}}^{t}\frac{d{\bar{\mu}}}{\bar{\mu}}
\gamma (\alpha _{s}({\bar{\mu}}^2))\right]\;,
\nonumber \\
S_{K}(t)&=&\exp\left[-s(x_{3}P_{3}^{-},b_{3})
-s((1-x_{3})P_{3}^{-},b_{3})
-2\int_{1/b_{3}}^{t}\frac{d{\bar{\mu}}}{\bar{\mu}}
\gamma (\alpha_{s}({\bar{\mu}}^2))\right]\;,
\label{sbk}
\end{eqnarray}
with the quark anomalous dimension $\gamma=-\alpha_s/\pi$.
The variables $b_{1}$, $b_{2}$, and $b_{3}$, conjugate to the parton
transverse momenta $k_{1T}$, $k_{2T}$, and $k_{3T}$, represent the
transverse extents of the $B$, $\phi$ and $K$ mesons, respectively.
The expression for the exponent $s$ is referred to \cite{CS,BS,LS}.
The above Sudakov exponentials decrease fast in the large $b$ region, such
that the $B\to\phi K$ hard amplitudes remain sufficiently perturbative
in the end-point region.

Recently, the importance of threshold resummation \cite{S0,CT,L2} has been 
observed in exclusive $B$ meson decays \cite{L5}. As $x_3\to 0$ (to be 
precise, $x_3\sim O(\bar\Lambda/M_B)$) in Fig.~2(a), the internal 
$b$ quark, carrying the momentum $P_1-k_3$, becomes almost on-shell, 
indicating that the end-point singularity is associated with the $b$ quark. 
Additional soft divergences then appear at higher orders, and the double 
logarithm $\alpha_s\ln^2 x_3$ is produced from the loop correction to the
weak decay vertex. This double logarithm can be factored out of the hard
amplitude systematically, and its resummation introduces a Sudakov factor
$S_t(x_3)$ into PQCD factorization formulas \cite{L5}. Similarly, another
lowest-order diagram Fig.~2(b) with a hard gluon exchange between the
$d(s)$ quark and the spectator quark gives an amplitude proportional to
$1/(x_1^2 x_3)$. In the threshold region with $x_1\to 0$ (to be precise,
$x_1\sim O(\bar\Lambda^2/M_B^2)$), additional collinear divergences are
associated with the internal $d(s)$ quark carrying the momentum
$P_3-k_1$. The double logarithm $\alpha_s\ln^2 x_1$ is then produced from
the loop correction to the weak decay vertex. Resummation of this type of
double logarithms leads to the Sudakov factor $S_t(x_1)$.

The above formalism can be generalized to factorizable annihilation
diagrams easily. For the lowest-order diagram with the internal quark
carrying the momentum $P_2+k_3$, the end-point region corresponds to
$x_3\to 0$. Hence, threshold resummation of the double logarithm gives
the Sudakov factor $S_t(x_3)$. For the lowest-order diagram with the
internal quark carrying the momentum $P_3+k_2$, the end-point region
corresponds to $x_2\to 0$. Threshold resummation of the double logarithm
then gives the Sudakov factor $S_t(x_2)$. The Sudakov factor from
threshold resummation is universal, independent of flavors of internal
quarks, twists and topologies of hard amplitudes, and decay modes. To
simplify the analysis, we have proposed the parametrization \cite{TLS},
\begin{eqnarray}
S_t(x)=\frac{2^{1+2c}\Gamma(3/2+c)}{\sqrt{\pi}\Gamma(1+c)} [x(1-x)]^c\;.
\label{str}
\end{eqnarray}
with the parameter $c=0.3$. This parametrization, symmetric under the
interchange of $x$ and $1-x$, is convenient for evaluation of factorizable
annihilation amplitudes. It is obvious that threshold resummation
modifies the end-point behavior of the meson distribution amplitudes,
rendering them vanish faster at $x\to 0$.

Threshold resummation for nonfactorizable diagrams is weaker and
negligible. The reason is understood as follows. Consider the diagram
with a hard gluon exchange between the spectator quark and the $s$ quark
in the $\phi$ meson, in which the internal $s$ quark carries the
momentum $k_2-k_1+k_3$. To obtain additional infrared divergences,
{\it i.e.}, the double logarithms, the $s$ quark must be close to
mass shell. We then have the threshold region defined, for example, by
$k_1\sim O(\bar\Lambda^2/M_B)$, $k_2\sim O(\bar\Lambda^2/M_B)$, and
$k_3\sim O(M_B)$ simultaneously. That is, this region has more limited
phase space compared to that for factorizable amplitudes. Furthermore,
soft contribution to a pair of nonfactorizable diagrams cancels, such that
the end-point region is not important. Based on the above observations, we 
shall not include threshold resummation for nonfactorizable amplitudes.

$k_T$ and threshold resummations arise from different subprocesses in
PQCD factorization. They can be derived in perturbation theory, and are
not free parameters. Their combined effect suppresses the end-point
contributions, making PQCD evaluation of exclusive $B$ meson decays
reliable. If excluding the resummation effects, the PQCD predictions
for the $B\to K$ form factor are infrared divergent. If including only
$k_T$ resummation, the PQCD predictions are finite. However, the
two-parton twist-3 contributions are still huge, so that the $B\to K$
form factor has an unreasonably large value $F^{BK}\sim 0.57$ at maximal
recoil. The reason is that the double logarithms $\alpha_s\ln^2 x$ have
not yet been organized. If including both resummations, we obtain the
reasonable result $F^{BK}\sim 0.35$. This study indicates the importance
of resummations in PQCD analyses of $B$ meson decays. In conclusion, if
the PQCD evaluation of the heavy-to-light form factors is performed
self-consistently, there exist no end-point singularities, and both
twist-2 and twist-3 contributions are well-behaved.
 
The mechanism of Sudakov suppression can be easily understood by regarding
a meson as a color dipole. In the region with vanishing $k_T$ and $x$,
the meson possesses a huge extent in the transverse and longitudinal
directions, respectively. That is, the meson carries a large color dipole.
At fast recoil, this large color dipole, strongly scattered during $B$
meson decays, tends to emit infinitely many real gluons. However, these
emissions are forbidden in an exclusive process with final-state particles
specified. As a consequence, contributions to the $B\to\pi, K$ form
factors from the kinematic region with vanishing $k_T$ and $x$ must be
highly suppressed.

\subsection{Form Factors}

We demonstrate that $k_T$ and threshold resummations must be taken into 
account in order to obtain reasonable results for the $B\to K, \pi$ form 
factors. In the PQCD approach these form factors are derived from the
diagrams with one hard gluon exchange as shown in Fig.~2. Soft
contribution from the diagram without any hard gluon is Sudakov suppressed
\cite{TLS}. For a rigorous justification of this statement in QCD sum
rules, refer to \cite{SR}. The two form factors $F^{BK}_{+,0}(q^2)$ are
defined by
\begin{eqnarray}
\langle K(P_3)|{\bar b}(0)\gamma_\mu s(0)|B(P_1)\rangle
&=& F_+^{BK}(q^2)\left[(P_1+P_3)_\mu-\frac{M_B^2-M_K^2}{q^2}q_\mu\right]
\nonumber\\
& &+ F_0^{BK}(q^2)\frac{M_B^2-M_K^2}{q^2}q_\mu\;,
\label{bkd}
\end{eqnarray}
where $q=P_1-P_3$ is the outgoing lepton-pair momentum. $F_+^{BK}$ and
$F_0^{BK}$ at maximal recoil are, quoted from Eq.~(\ref{Fe}) below,
written as
\begin{eqnarray}
F_{+,0}^{BK}(0) &=&8\pi
C_{F}M_{B}^{2}\int_{0}^{1}dx_{1}dx_{3}\int_{0}^{\infty}
b_{1}db_{1}b_{3}db_{3}\Phi _{B}\left( x_{1},b_{1}\right)  
\nonumber \\
&&\times \bigg\{ \left[
\left(1+x_{3}\right) \Phi _{K}\left( x_{3}\right)
+\left( 1-2x_{3}\right) r_{K}\left( \Phi _{K}^{p}
\left(x_{3}\right) +\Phi _{K}^{\sigma }\left( x_{3}\right) \right) 
\right] 
\nonumber \\
&&\times \alpha_s(t_e^{(1)}) S_B(t_e^{(1)}) S_K(t_e^{(1)}) 
h_{e}\left(x_{1},x_{3},b_{1},b_{3}\right)  \nonumber \\
& &+ 2r_{K}\Phi _{K}^{p}\left( x_{3}\right) 
\alpha_s(t_e^{(2)}) S_B(t_e^{(2)}) S_K(t_e^{(2)})
h_{e}\left( x_{3},x_{1},b_{3},b_{1}\right)
\bigg\} \;,  
\label{FormBK}
\end{eqnarray}
with $C_F=4/3$ being a color factor. The hard function
$h_e(x_1,x_3,b_1,b_3)$, referred to Eq.~(\ref{he}), contains the
threshold resummation factor $S_t(x_3)$. The hard scales $t_e$ are
defined in Eq.~(\ref{te12}). The expression for the $B\to\pi$ form
factor $F^{B\pi}(q^2)$ is similar to Eq.~(\ref{FormBK}). 

In Eq.~(\ref{FormBK}) we have included the complete two-parton twist-3
distribution amplitudes associated with the pseudo-scalar and
pseudo-tensor structures of the kaon. We adopt the mass $m_0=1.7$ GeV
\cite{Keum:2001ph,Keum:2001wi}, the $B$ meson distribution amplitude
$\Phi_B$ proposed in \cite{Keum:2001ph,Keum:2001wi} (see Eq.~(\ref{bw})),
and the kaon distribution amplitudes $\Phi_K$, $\Phi_K^p$, and
$\Phi_K^\sigma$ derived from QCD sum rules \cite{PB2}
(see Eqs.~(\ref{kk})-(\ref{ks})). Since $\Phi_B$ is still not well
determined, we consider the variation of its shape parameter within
0.36 GeV $< \omega_B <0.44$ GeV \cite{BCP4-K} around its
central value $\omega_B=0.4$ GeV. Turning off $k_T$ and threshold
resummations and fixing $\alpha_s$, the $B\to \pi, K$ form factors are
divergent and not calculable. With $k_T$ resummation, the results become
finite as shown in Table I, but still much larger than the expected
ones 0.3-0.4. Further including the threshold resummation effect, we
obtain the reasonable values $F^{BK}(0)\sim 0.35\pm 0.06$ and
$F^{B\pi}(0)=0.30 \pm 0.04$. These ranges of the form factors have been
usually adopted as model inputs in the literature, and are consistent
with the results from lattice calculations \cite{aoki,UKQCD,DB}
extrapolated to the small $q^2$ region and from light-cone QCD sum rules
\cite{KR,PB3}.

We also present our results of the time-like form factors $F^{\phi K}$,
which govern the factorizable annihilation amplitudes. The factorization
formulas are, quoted from Eqs.~(\ref{Fa4}) and (\ref{Fa6}), given by
\begin{eqnarray}
F_{(V-A)}^{\phi K}&=&8\pi
C_{F}M_{B}^{2}\int_{0}^{1}dx_{2}dx_{3}\int_{0}^{\infty}
b_{2}db_{2}b_{3}db_{3}  
\nonumber \\
& &\times \bigg\{ \left[ \left( 1-x_{3}\right) 
\Phi_{\phi} ( x_{2}) \Phi_{K}\left( x_{3}\right)
 +2r_K r_{\phi}\Phi _{\phi}^{s}\left( x_{2}\right) \left(
(2-x_{3})\Phi_{K}^{p}\left( x_{3}\right) +x_{3}\Phi_{K}^{\sigma }\left(
x_{3}\right) \right) \right] 
\nonumber \\
& &\times \alpha_s(t_a^{(1)}) S_{\phi}(t_a^{(1)}) S_K(t_a^{(1)}) 
h_{a}\left(x_{2},1-x_{3},b_{2},b_{3}\right)   
\nonumber \\
& &-\left[ x_{2} \Phi_{\phi}\left( x_{2}\right) 
\Phi_{K}\left( x_{3}\right)  +2r_K r_{\phi}
\left((1+x_{2})\Phi_{\phi}^{s}\left( x_{2}\right)
-(1-x_{2})\Phi_{\phi}^{t}\left( x_{2}\right)
\right) \Phi_{K}^{p}\left( x_{3}\right) \right]   
\nonumber \\
& &\times \alpha_s(t_a^{(2)}) S_{\phi}(t_a^{(2)}) S_K(t_a^{(2)}) 
h_{a}\left(1-x_{3},x_{2},b_{3},b_{2}\right) \bigg\}\;.  
\label{phi-k-1} \\
\cr
F_{(V+A)}^{\phi K} &=&-8\pi
C_{F}M_{B}^{2}\int_{0}^{1}dx_{2}dx_{3}\int_{0}^{\infty}
b_{2}db_{2}b_{3}db_{3}
\nonumber \\
& &\times \bigg\{ \left[ 2 r_{K} ( 1-x_{3}) \Phi _{\phi}(x_{2}) 
\left( \Phi _{K}^{p} ( x_{3}) + \Phi_{K}^{\sigma }(x_{3}) \right) 
+4r_{\phi }\Phi _{\phi}^{s}\left( x_{2}\right) \Phi_{K}
\left( x_{3}\right) \right] 
\nonumber \\
& &\times \alpha_s(t_a^{(1)}) S_{\phi}(t_a^{(1)}) S_K(t_a^{(1)}) 
h_{a}\left(x_{2},1-x_{3},b_{2},b_{3}\right)  
\nonumber \\
& &+\left[ 4r_{K}\Phi_{\phi}(x_2)\Phi _{K}^{p}(x_3)+2 x_{2} r_{\phi}
\left(\Phi_{\phi}^{s}( x_{2})-\Phi _{\phi}^{t}\left( x_{2}\right)
\right)  \Phi_{K}( x_{3})   \right]
\nonumber \\
& & \times  \alpha_s(t_a^{(2)}) S_{\phi}(t_a^{(2)}) S_K(t_a^{(2)}) 
h_{a}\left(1-x_{3},x_{2},b_{3},b_{2}\right) \bigg\}\;. 
\label{phi-k-2}
\end{eqnarray}
The hard function $h_a(x_2,x_3,b_2,b_3)$, referred to Eq.~(\ref{ha}),
contains the threshold resummation factor $S_t(x_3)$. The hard scales
$t_a$ and the $\phi$ meson distribution amplitudes $\Phi_\phi$,
$\Phi_\phi^t$ and $\Phi_\phi^s$ are defined in Eq.~(\ref{et}) and in
Eqs.~(\ref{phi2})-(\ref{phi3s}), respectively. It is obvious that
$F_{(V-A)}^{\phi K}$ contains both logarithmic and linear divergences,
and $F_{(V+A)}^{\phi K}$ contains logarithmic divergences \cite{CY,DYZ},
if the Sudakov factors are excluded. To include annihilation
contributions in the QCD factorization approach, several arbitrary
complex infrared cutoffs must be introduced to regulate the above
end-point singularities \cite{BBNS3}. These cutoffs are process-dependent.
Hence, the PQCD approach with the Sudakov effects has a better control on
annihilation contributions.
 
We obtain $F^{\phi K}_{(V-A)} = (1.78 + 0.63 i) \times 10^{-2}$ and
$F^{\phi K}_{(V+A)} = (-3.48 + 13.52 i) \times 10^{-2}$ for
$\omega_B=0.4$ GeV and $m_0=1.7$ GeV. The smaller value of
$F^{\phi K}_{(V-A)}$ is due to mechanism similar to helicity suppression:
note the partial cancellation between the two terms in Eq.~(\ref{phi-k-1}).
$F^{\phi K}_{(V+A)}$ is mainly imaginary, whose reason will become clear
after we explain the power behavior of the various decay amplitudes in
the next section. The time-like form factors are independent of the shape
parameter $\omega_B$. The change of the $B\to\pi, K$ form factors and of
the time-like form factors with the chiral enhancement factor $m_0$ is
displayed in Table II.

\section{POWER COUNTING AND DECAY AMPLITUDES}

We discuss the power counting rules in the presence of the Sudakov
effects and present the factorization formulas for the $B\to \phi K$
decays. The effective Hamiltonian for the flavor-changing $b\to s$
transition is given by
\begin{equation}
H_{{\rm eff}}={\frac{G_{F}}{\sqrt{2}}}\sum_{q=u,c}V_{q}\left[ C_{1}(\mu)
O_{1}^{(q)}(\mu )+C_{2}(\mu )O_{2}^{(q)}(\mu )+\sum_{i=3}^{10}C_{i}(\mu)
O_{i}(\mu )\right] \;,  \label{hbk}
\end{equation}
with the Cabibbo-Kobayashi-Maskawa (CKM) matrix elements
$V_{q}=V_{qs}^{*}V_{qb}$ and the operators 
\begin{eqnarray}
&&O_{1}^{(q)}=(\bar{s}_{i}q_{j})_{V-A}(\bar{q}_{j}b_{i})_{V-A}\;,\;\;\;\;\;
\;\;\;O_{2}^{(q)}=(\bar{s}_{i}q_{i})_{V-A}(\bar{q}_{j}b_{j})_{V-A}\;, 
\nonumber \\
&&O_{3}=(\bar{s}_{i}b_{i})_{V-A}\sum_{q}(\bar{q}_{j}q_{j})_{V-A}\;,\;\;\;
\;O_{4}=(\bar{s}_{i}b_{j})_{V-A}\sum_{q}(\bar{q}_{j}q_{i})_{V-A}\;, 
\nonumber \\
&&O_{5}=(\bar{s}_{i}b_{i})_{V-A}\sum_{q}(\bar{q}_{j}q_{j})_{V+A}\;,\;\;\;
\;O_{6}=(\bar{s}_{i}b_{j})_{V-A}\sum_{q}(\bar{q}_{j}q_{i})_{V+A}\;, 
\nonumber \\
&&O_{7}=\frac{3}{2}(\bar{s}_{i}b_{i})_{V-A}\sum_{q}e_{q} (\bar{q}%
_{j}q_{j})_{V+A}\;,\;\;O_{8}=\frac{3}{2}(\bar{s}_{i}b_{j})_{V-A}
\sum_{q}e_{q}(\bar{q}_{j}q_{i})_{V+A}\;,  \nonumber \\
&&O_{9}=\frac{3}{2}(\bar{s}_{i}b_{i})_{V-A}\sum_{q}e_{q} (\bar{q}%
_{j}q_{j})_{V-A}\;,\;\;O_{10}=\frac{3}{2}(\bar{s}_{i}b_{j})_{V-A}
\sum_{q}e_{q}(\bar{q}_{j}q_{i})_{V-A}\;,
\end{eqnarray}
$i$ and $j$ being the color indices. Using the unitarity condition, the
CKM matrix elements for the penguin operators $O_{3}$-$O_{10}$ can also
be expressed as $V_{u}+V_{c}=-V_{t}$. The unitarity angle $\phi_{3}$ is
defined via 
\begin{equation}
V_{ub}=|V_{ub}|\exp (-i\phi _{3})\;.
\end{equation}
Here we adopt the Wolfenstein parametrization for the CKM matrix upto
$O(\lambda^{3})$: 
\begin{eqnarray}
\left(\matrix{V_{ud} & V_{us} & V_{ub} \cr V_{cd} & V_{cs} & V_{cb} \cr
V_{td} & V_{ts} & V_{tb} \cr}\right) =\left(\matrix{ 1 - \lambda^2/2 &
\lambda & A \lambda^3(\rho - i \eta)\cr - \lambda & 1 - \lambda^2/ 2 & A
\lambda^2\cr A \lambda^3(1-\rho-i\eta) & -A \lambda^2 & 1 \cr}\right)\;.
\end{eqnarray}
A recent analysis of quark-mixing matrix yields \cite{LEP} 
\begin{eqnarray}
\lambda &=& 0.2196 \pm 0.0023\;,  \nonumber \\
A &=& 0.819 \pm 0.035\;,  \nonumber \\
R_b &\equiv&\sqrt{{\rho}^2 + {\eta}^2} = 0.41 \pm 0.07\;.
\end{eqnarray}

The hard amplitudes contain factorizable diagrams, where hard gluons
attach the valence quarks in the same meson, and nonfactorizable
diagrams, where hard gluons attach the valence quarks in different
mesons. The annihilation topology is also included, and classified into
factorizable and nonfactorizable ones according to the above definitions. 
The $B\to \phi K$ decay rates have the expressions
\begin{equation}
\Gamma =\frac{G_{F}^{2}M_{B}^{3}}{32\pi }|A|^{2}\;.
\label{dr1}
\end{equation}
The amplitudes for $B_{d}\rightarrow \phi \bar{K}^{0}$,
$\bar{B}_{d}\rightarrow \phi K^{0}$, $B^{+}\rightarrow \phi K^{+}$ and
$B^{-}\rightarrow \phi K^{-}$ are written, respectively, as
\begin{eqnarray}
A &=&f_{\phi }V_{t}^{*}F_{e}^{P(s)}+V_{t}^{*}{\cal M}_{e}^{P(s)}
+f_{B}V_{t}^{*}F_{a}^{P(d)}+V_{t}^{*}{\cal M}_{a}^{P(d)}\;,
\label{namp1} \\
\bar{A} &=&f_{\phi }V_{t}F_{e}^{P(s)}+V_{t}{\cal M}_{e}^{P(s)}
+f_{B}V_{t}F_{a}^{P(d)}+V_{t}{\cal M}_{a}^{P(d)}\;,
\label{namp2}\\
A^{+} &=&f_{\phi }V_{t}^{*}F_{e}^{P(s)}+V_{t}^{*}{\cal M}_{e}^{P(s)}
+f_{B}V_{t}^{*}F_{a}^{P(u)}+V_{t}^{*}{\cal M}_{a}^{P(u)}
-f_{B}V_{u}^{*}F_{a}-V_{u}^{*}{\cal M}_{a}\;,
\label{camp1} \\
A^{-} &=&f_{\phi }V_{t}F_{e}^{P(s)}+V_{t}{\cal M}_{e}^{P(s)}
+f_{B}V_{t}F_{a}^{P(u)}+V_{t}{\cal M}_{a}^{P(u)}
-f_{B}V_{u}F_{a}-V_{u}{\cal M}_{a}\;.
\label{camp2}
\end{eqnarray}
In the above expressions $F$ (${\cal M}$) denote factorizable
(nonfactorizable) amplitudes, the subscripts $e$ ($a$) denote the
emission (annihilaiton) diagrams, and the superscripts $P(q)$ denote
amplitudes from the penguin operators involving the $q$-$\bar q$ quark 
pair, and $f_B$ ($f_\phi$) is the $B$ ($\phi$) meson decay constant.

\subsection{Power Counting}

Before presenting the factorization formulas of the above amplitudes, we
discuss their power behavior in $1/M_B$. The spectator quark in the $B$
meson, forming a soft cloud around the heavy $b$ quark, carries momentum
of $O(\bar\Lambda)$. The spectator quark on the kaon side carries
momentum of $O(M_B)$ in order to form the fast-moving kaon with the $s$
quark produced in the $b$ quark decay. These dramatic different orders of
magnitude in momenta explain why a hard gluon is necessary. Based on this
argument, the hard gluon is off-shell by order of $\bar\Lambda M_B$. This
special scale, characterizing heavy-to-light decays, is essential for
developing the PQCD formalism of exclusive $B$ meson decays. Below we
shall explicitly show how to construct this power behavior, and argue
that all the topologies of diagrams should be taken into account in the
leading-power PQCD analysis.

We start with the twist-2 contribution to the factorizable emission
amplitude, which contains the integrand (see Eq.~(\ref{FormBK})),
\begin{eqnarray}
I^{e2}=\frac{x_3M_B^2}{[x_1x_3M_B^2+O(\bar\Lambda M_B)][x_3M_B^2+
O(\bar\Lambda M_B)]}\;,
\end{eqnarray}
where the factor $x_3$ in the numerator comes from the twist-2 kaon
distribution amplitude $\Phi_K(x_3)$, the first and second factors in 
the denominator come from the virtual gluon and quark propagators,
respectively. The terms of $O(\bar\Lambda M_B)$ represent the order of
magnitude of parton transverse momenta $k_T^2$ under Sudakov suppression.
The momentum fraction $x_1$ is assumed to be of $O(\bar\Lambda/M_B)$.
Here we concentrate on the important end-point region with
$x_3\sim O(\bar\Lambda/M_B)$. It is trivial to find that in the
end-point region $I^{e2}$ behaves like
\begin{eqnarray}
I^{e2}\sim \frac{1}{\bar\Lambda M_B}\;,
\label{if2}
\end{eqnarray}
as argued above.

We then consider the first twist-3 term in Eq.~(\ref{FormBK}),
\begin{eqnarray}
I^{e3}=\frac{r_KM_B^2}{[x_1x_3M_B^2+O(\bar\Lambda M_B)][x_3M_B^2+
O(\bar\Lambda M_B)]}\;.
\end{eqnarray}
For small $x_3\sim O(\bar\Lambda/M_B)$, we have the power law,
\begin{eqnarray}
I^{e3}\sim \frac{r_K}{\bar\Lambda^2}= \frac{m_0}{\bar\Lambda}
\frac{1}{\bar\Lambda M_B}\;.
\end{eqnarray}
Hence, the twist-3 contribution is not power-suppressed compared to the
twist-2 one in the $M_B\to\infty$ limit \cite{L6,TLS}, though the
two-parton twist-3 distribution amplitudes are proportional to the ratio
$r_K=m_0/M_B$. The presence of the potential linear divergence
in the twist-3 contribution modifies the naive power counting rules from 
twist expansion of meson distribution amplitudes. The power behavior of
the other twist-3 term in Eq.~(\ref{FormBK}) is the same.

A folklore for annihilation contributions is that they are suppressed
by a power of the small ratio $f_B/M_B$, and negligible compared to
emission contributions. The annihilation amplitudes from the operators
$O_{1,2,3,4}$ with the structure $(V-A)(V-A)$ are small because of
mechanism of helicity suppression. This argument applies exactly to the
$B\to\pi\pi$ decays, and partially to the $B\to K\pi$ and $\phi K$
decays, since the kaon distribution amplitudes are not symmetric in the
momentum fraction $x$, and the two final-state mesons are not identical.
Those from the operators $O_{5,6}$ with the structure $(S-P)(S+P)$
survive under helicity suppression \cite{Keum:2001ph,Keum:2001wi}. Below
we shall argue that the annihilation amplitudes from $O_{5,6}$ are
proportional to $2r_K$, which is in fact of $O(1)$ for $M_B\sim 5$ GeV.
That is, annihilation contributions vanish in the heavy quark limit
$m_b\to \infty$, but are important for the physical mass $m_b$.
 
Referred to Eq.~(\ref{phi-k-1}), the integrand for the twist-2 
annihilation amplitudes from $O_{1,2,3,4}$ is written as,
\begin{eqnarray}
I_{(V-A)}^{a2}=\frac{x_2x_3^2M_B^2}{[x_2x_3M_B^2-O(\bar\Lambda M_B)
+i\epsilon][x_3M_B^2-O(\bar\Lambda M_B)+i\epsilon]}\;,
\label{ia2}
\end{eqnarray}
where the factor $x_2$ in the numerator comes from $\Phi_\phi(x_2)$ and
$x_3^2$ comes from $x_3\Phi_K(x_3)$. Here we have interchanged $x_3$ and
$1-x_3$ for the convenience of discussion. In the region with
$x_3\sim O(\bar{\Lambda}/M_B)$ and with arbitrary $x_2$, we obtain the
power laws of the real and imaginary parts of Eq.~(\ref{ia2}),
\begin{eqnarray}
Re(I_{(V-A)}^{a2})\sim \frac{1}{M_B^2}\;,\;\;\;\;
Im(I_{(V-A)}^{a2})\sim \frac{1}{M_B^2}\;.
\end{eqnarray}
Hence, the twist-2 annihilation contributions from $O_{1,2,3,4}$ are
negligible.

The integrand for the $O(r^2)$ terms in Eq.~(\ref{phi-k-1}) is written as,
\begin{eqnarray}
I_{(V-A)}^{a4}=\frac{2r_Kr_\phi M_B^2}{[x_2x_3M_B^2-O(\bar\Lambda M_B)
+i\epsilon][x_3M_B^2-O(\bar\Lambda M_B)+i\epsilon]}\;.
\label{ia4}
\end{eqnarray}
We express the quark propagator as
\begin{eqnarray}
\frac{1}{x_3M_B^2-O(\bar\Lambda M_B)+i\epsilon}=
\frac{P}{x_3M_B^2-O(\bar\Lambda M_B)}-\frac{i\pi}{M_B^2}
\delta\left(x_3-O\left(\frac{\bar\Lambda}{M_B}\right)\right)\;,
\label{pvm}
\end{eqnarray}
where $P$ denotes the principle-value prescription. The gluon propagator
is expressed in a similar way. The real part then behaves like
\begin{eqnarray}
Re(I_{(V-A)}^{a4})&=&2r_K r_\phi M_B^2\left[
\frac{P}{[x_2x_3M_B^2-O(\bar\Lambda M_B)]}
\frac{P}{[x_3M_B^2-O(\bar\Lambda M_B)]}\right.
\nonumber\\
& &\left.-\frac{\pi^2}{M_B^4}
\delta\left(x_2x_3-O\left(\frac{\bar\Lambda}{M_B}\right)\right)
\delta\left(x_3-O\left(\frac{\bar\Lambda}{M_B}\right)\right)\right]\;,
\nonumber\\
&\sim& 2r_Kr_\phi\left(\frac{1}{\bar\Lambda^2}
-\frac{\pi^2}{\bar\Lambda M_B}\right)
= 2r_K\frac{M_\phi}{\bar\Lambda}\frac{1}{\bar\Lambda M_B}
\left(1-\frac{\pi^2\bar\Lambda}{M_B}\right)\;.
\label{ia4r}
\end{eqnarray}
The imaginary part of Eq.~(\ref{ia4}) behaves like
\begin{eqnarray}
Im(I_{(V-A)}^{a4})&=&-2\pi r_K r_\phi\left[
\frac{P}{[x_2x_3M_B^2-O(\bar\Lambda M_B)]}
\delta\left(x_3-O\left(\frac{\bar\Lambda}{M_B}\right)\right)\right.
\nonumber\\
& &\left.+\frac{P}{[x_3M_B^2-O(\bar\Lambda M_B)]}
\delta\left(x_2x_3-O\left(\frac{\bar\Lambda}{M_B}\right)\right)
\right]\;,
\nonumber\\
&\sim& -2r_K r_\phi\frac{2\pi}{\bar\Lambda M_B}
= -2r_K \frac{M_\phi}{\bar\Lambda}\frac{2\pi}{M_B^2}\;.
\label{ia4i}
\end{eqnarray}
For the above estimates the end point $x_3\sim O(\bar\Lambda/M_B)$ with
arbitrary $x_2$ corresponds to the important region. Note that the ratio
$M_\phi/\bar\Lambda$ is of $O(1)$. Compared to Eq.~(\ref{if2}), the first
term in Eq.~(\ref{ia4r}), multiplied by a chiral factor $2r_K\sim O(1)$,
is not down by the small ratio $f_B/M_B$. The second term in
Eq.~(\ref{ia4r}) and the imaginary part, though scaling like $1/M_B^2$,
are enhanced by the factors $\pi^2$ and $2\pi$, respectively. However,
the mechanism of helicity suppression renders these annihilation
contributions turn out to be small ($\sim 1/M_B^2$) as shown in Table II.

Referred to Eq.~(\ref{phi-k-2}), the general integrand for the twist-3
annihilation amplitudes from $O_{5,6}$ is given by,
\begin{eqnarray}
I_{(V+A)}^{a3}=\frac{2r_K x_2x_3M_B^2}{[x_2x_3M_B^2-O(\bar\Lambda M_B)
+i\epsilon][x_3M_B^2-O(\bar\Lambda M_B)+i\epsilon]}\;,
\label{ia3}
\end{eqnarray}
By means of a similar argument, the real and imaginary parts behave,
respectively, like
\begin{eqnarray}
Re(I_{(V+A)}^{a3}) & \sim & 2r_K\frac{1}{\bar\Lambda M_B}
\left(1-\frac{\pi^2\bar\Lambda}{M_B}\right)\;,
\label{ia3r}\\
Im(I_{(V+A)}^{a3})& \sim & -2r_K\frac{2\pi}{M_B^2}\;,
\label{ia3i}
\end{eqnarray}
It is observed that the twist-3 amplitudes in Eq.~(\ref{phi-k-2})
possess the same power law as the $r^2$ terms in Eq.~(\ref{phi-k-1}).
The difference is only the $O(1)$ ratio $M_\phi/\bar\Lambda$. As stated
before, the potential linear divergence in the $r^2$ terms alters the
naive power counting rules from twist expansion.

Because of $O(M_B/\bar\Lambda)\sim \pi^2$ for $M_B\sim 5$ GeV, the two
terms in Eq.~(\ref{ia3r}) almost cancel each other. The imaginary part
in Eq.~(\ref{ia3i}) is enhanced by the factor $2\pi$. There is no
helicity suppression in this case. It is then understood that the
annihilation amplitudes from $O_{5,6}$ are mainly imaginary, and their
magnitude is few times smaller than the factorizable emission one as
exhibited in Table II. If the mass $M_B$ changes, the relative importance
of the real and imaginary parts will change. For example, as $M_B$
increases, the cancellation in Eq.~(\ref{ia3r}) is not exact, such that
the real part becomes larger. On the other hand, the imaginary part in
Eq.~(\ref{ia3i}) decreases. It has been checked that for $M_B=10$ GeV,
$Re(F_{(V+A)}^{\phi K})$ is of the same order as $Im(F_{(V+A)}^{\phi K})$. 
We conclude that the smallness of $Re(F_{(V+A)}^{\phi K})$ in Table II is
due to the special value of the $B$ meson mass $M_B\sim 5$ GeV. For
general $M_B$, it is more appropriate to state that annihilation
amplitudes from $O_{5,6}$, without distinguishing their real and imaginary
parts, scale like $2r_K/(\bar\Lambda M_B)$ according to Eq.~(\ref{ia3r}).

The above reasoning is applicable to nonfactorizable amplitudes. It can
be found, referred to Eqs.~(\ref{Me4})-(\ref{Me6}) and to the asymptotic
behavior of the meson distribution amplitudes, that the twist-2 term of
each nonfactorizable diagram scales like $1/(\bar\Lambda M_B)$. However,
because of the soft cancellation between a pair of nonfactorizable
diagrams in the end-point region of $x_3$, the sum of twist-2 terms turns
out to scale like $1/M_B^2$. The twist-3 and $O(r^2)$ terms in each
nonfactorizable diagram scale like
\begin{eqnarray}
\frac{r_K}{\bar\Lambda M_B}\;,\;\;\;\;
\frac{r_K r_\phi}{\bar\Lambda^2}=r_K
\frac{M_\phi}{\bar\Lambda}\frac{1}{\bar\Lambda M_B}\;,
\end{eqnarray}
respectively. The cancellation between a pair of nonfactorizable diagrams
modifies the above power behaviors into
\begin{eqnarray}
\frac{r_K}{M_B^2}\;,\;\;\;\;
\frac{r_K r_\phi}{\bar\Lambda M_B}=r_K
\frac{M_\phi}{\bar\Lambda}\frac{1}{M_B^2}\;,
\end{eqnarray}
respectively. For the nonfactorizable annihilation amplitudes (see
Eqs.~(\ref{Ma4}) and (\ref{Ma6})), the soft cancellation does not exist,
since the $B$ meson is a heavy-light system. However, it can be shown
that the twist-2, twist-3 and $O(r^2)$ terms in each nonfactorizable
annihilation diagram possess the power behaviors,
\begin{eqnarray}
\frac{1}{M_B^2}\;,\;\;\;\;
\frac{r_K}{M_B^2}\;,\;\;\;\;
r_K\frac{M_\phi}{\bar\Lambda}\frac{1}{M_B^2}\;,
\end{eqnarray}
respectively.

We emphasize that it is more appropriate to count the power of each
individual diagram, instead of the power of sum of diagrams. In some
cases, factorizable contributions are suppressed by a vanishing Wilson
coefficient, so that nonfactorizable contributions become dominant. For
example, factorizable internal-$W$ emisson contributions are strongly
suppressed by the Wilson coefficient $a_2$ in the $B\to J/\psi K^{(*)}$
decays \cite{YL}. In some cases, such as the $B\to D\pi$ decays, there 
is no soft cancellation between a pair of nonfactorizable diagrams, and 
nonfactorizable contributions are significant \cite{YL}. In summary, we
derive the relative importance of the various topologies of amplitudes
given in Eq.~(\ref{rim}).  The annihilation and nonfactorizable 
amplitudes are indeed negligible compared to the factorizable emission 
ones in the heavy quark limit. However, for the physical mass 
$M_B\sim 5$ GeV, the annihilation contributions should be included.


\subsection{Factorization Formulas}

Below we calculate the hard amplitudes for the emission and annihilation
topologies, whcih have been obtained independently in \cite{Mi}. The 
factorizable penguin contribution is written as 
\begin{eqnarray}
F_{e}^{P\left( s\right) } &=&-8\pi
C_{F}M_{B}^{2}\int_{0}^{1}dx_{1}dx_{3}\int_{0}^{\infty
}b_{1}db_{1}b_{3}db_{3}\Phi _{B}\left( x_{1},b_{1}\right)  
\nonumber \\
&&\times \bigg\{ \left[\left(1+x_{3}\right) \Phi _{K}\left( x_{3}\right)
+r_K ( 1-2x_{3}) \left( \Phi _{K}^{p}\left(
x_{3}\right) +\Phi _{K}^{\sigma }\left( x_{3}\right) \right)
\right] 
\nonumber \\
&&\times E_{e}^{(s)}\left( t_{e}^{\left( 1\right) }\right) h_{e}\left(
x_{1},x_{3},b_{1},b_{3}\right)  
\nonumber \\
&&+ 2r_{K}\Phi _{K}^{p}\left( x_{3}\right)  E_{e}^{(s)}\left(
t_{e}^{\left( 2\right) }\right) h_{e}\left( x_{3},x_{1},b_{3},b_{1}\right)
\bigg\} \;.
\label{Fe}
\end{eqnarray}
The factorizable annihilation contribution is given by
\begin{equation}
F_{a}^{P(q)} = F_{a4}^{P\left( q\right) }+F_{a6}^{P(q)}\;,
\label{Fa}
\end{equation}
with 
\begin{eqnarray}
F_{a4}^{P\left( q\right) } &=&8\pi
C_{F}M_{B}^{2}\int_{0}^{1}dx_{2}dx_{3}\int_{0}^{\infty
}b_{2}db_{2}b_{3}db_{3}  
\nonumber \\
&&\times \bigg\{ \left[ \left( 1-x_{3}\right) 
\Phi_{\phi} ( x_{2}) \Phi_{K}\left( x_{3}\right)\right. 
\nonumber \\
&&\left. +2r_{K}r_{\phi}\Phi _{\phi}^{s}\left( x_{2}\right) \left(
(2-x_{3})\Phi_{K}^{p}\left( x_{3}\right) +x_{3}\Phi_{K}^{\sigma }\left(
x_{3}\right) \right) \right]   
\nonumber \\
&&\times E_{a4}^{(q)}\left( t_{a}^{\left( 1\right) }\right) h_{a}\left(
x_{2},1-x_{3},b_{2},b_{3}\right)   
\nonumber \\
&&-\left[ x_{2} \Phi_{\phi}\left( x_{2}\right) 
\Phi_{K}\left( x_{3}\right) \right.   
\nonumber \\
&&\left. +2r_{K}r_{\phi} \left((1+x_{2})\Phi_{\phi}^{s}\left(x_{2}\right)
-(1-x_{2})\Phi_{\phi}^{t}\left( x_{2}\right)
\right) \Phi_{K}^{p}\left( x_{3}\right) \right]
\nonumber \\
&&\times E_{a4}^{(q)}\left( t_{a}^{\left( 2\right) }\right)
h_{a}\left(1-x_{3},x_{2},b_{3},b_{2}\right) \bigg\} \;,
\label{Fa4} \\
\cr
F_{a6}^{P\left( q\right) } &=&-8\pi
C_{F}M_{B}^{2}\int_{0}^{1}dx_{2}dx_{3}\int_{0}^{\infty}
b_{2}db_{2}b_{3}db_{3}  
\nonumber \\
&&\times \bigg\{ \left[ 2 r_{K} ( 1-x_{3}) \Phi _{\phi}(x_{2}) 
\left( \Phi _{K}^{p} ( x_{3}) + \Phi_{K}^{\sigma }(x_{3}) \right)
+4r_{\phi }\Phi _{\phi}^{s}\left( x_{2}\right) \Phi_{K}
\left( x_{3}\right) \right] 
\nonumber \\
&&\times E_{a6}^{(q)}\left( t_{a}^{\left( 1\right) }\right) h_{a}\left(
x_{2},1-x_{3},b_{2},b_{3}\right)  
\nonumber \\
&&+\left[ 4r_{K}\Phi_{\phi}(x_2)\Phi _{K}^{p}(x_3)+2 x_{2} r_{\phi}
\left(\Phi_{\phi}^{s}( x_{2})-\Phi _{\phi}^{t}\left( x_{2}\right)
\right)  \Phi_{K}( x_{3})   \right]
\nonumber \\
&& \times E_{a6}^{(q)}\left( t_{a}^{\left( 2\right) }\right)
h_{a}\left(1-x_{3},x_{2},b_{3},b_{2}\right) \bigg\}\;,
\label{Fa6}
\end{eqnarray}
for the light quarks $q=u$ and $d$. $F_{a}$ in Eqs.~({\ref{camp1}) and
(\ref{camp2}) are the same as $F_{a4}^{P(u)}$ but with the Wilson
coefficient $a_{4}^{(u)}$ replaced by $a_{2}$. 

The factors $E(t)$ contain the evolution from the $W$ boson mass to the
hard scales $t$ in the Wilson coefficients $a(t)$, and from $t$ to the
factorization scale $1/b$ in the Sudakov factors $S(t)$:
\begin{eqnarray}
E_{e}^{(q)}\left( t\right) &=&\alpha _{s}\left( t\right) a_{e}^{(q)}(t)
S_{B}\left( t\right)S_{K}\left( t\right)\;,
\nonumber \\
E_{ai}^{(q)}\left( t\right) &=&\alpha _{s}\left( t\right) a_{i}^{(q)}(t)
S_{\Phi }(t)S_{K}(t) \;.
\label{Eea}
\end{eqnarray}
The hard functions $h$'s are 
\begin{eqnarray}
h_{e}(x_{1},x_{3},b_{1},b_{3}) &=&K_{0}\left( \sqrt{x_{1}x_{3}}
M_{B}b_{1}\right)S_t(x_3)
\nonumber \\
&&\times \left[ \theta (b_{1}-b_{3})K_{0}\left( \sqrt{x_{3}}
M_{B}b_{1}\right) I_{0}\left( \sqrt{x_{3}}M_{B}b_{3}\right) \right. 
\nonumber \\
&&\left. +\theta (b_{3}-b_{1})K_{0}\left( \sqrt{x_{3}}M_{B}b_{3}\right)
I_{0}\left( \sqrt{x_{3}}M_{B}b_{1}\right) \right] \;,
\label{he} \\
h_{a}(x_{2},x_{3},b_{2},b_{3}) &=&\left( \frac{i\pi }{2}\right)^{2}
H_{0}^{(1)}\left( \sqrt{x_{2}x_{3}}M_{B}b_{2}\right)S_t(x_3)
\nonumber \\
&&\times \left[ \theta (b_{2}-b_{3})H_{0}^{(1)}\left( \sqrt{x_{3}}
M_{B}b_{2}\right) J_{0}\left( \sqrt{x_{3}}M_{B}b_{3}\right) \right. 
\nonumber \\
&&\left. +\theta (b_{3}-b_{2})H_{0}^{(1)}\left( \sqrt{x_{3}}
M_{B}b_{3}\right) J_{0}\left( \sqrt{x_{3}}M_{B}b_{2}\right) \right] \;,
\label{ha}
\end{eqnarray}
where $S_{t}(x)$ is the evolution function from threshold resummation
discussed in Sec.~II, and $K_0, I_0, H_0$ and $J_0$ are the Bessel
functions.

The hard scales $t$ are chosen as the maxima of the virtualities of
internal particles involved in the hard amplitudes, including $1/b_{i}$: 
\begin{eqnarray}
t_{e}^{(1)} &=&{\rm max}(\sqrt{x_{3}}M_{B},1/b_{1},1/b_{3})\;,
\nonumber \\
t_{e}^{(2)} &=&{\rm max}(\sqrt{x_{1}}M_{B},1/b_{1},1/b_{3})\;,
\label{te12} \\
t_{a}^{(1)} &=&{\rm max}(\sqrt{1-x_{3}}M_{B},1/b_{2},1/b_{3})\;,
\nonumber\\
t_{a}^{(2)} &=&{\rm max}(\sqrt{x_{2}}M_{B},1/b_{2},1/b_{3})\;,
\label{et}
\end{eqnarray}
which decrease higher-order corrections \cite{CKL}. The Sudakov factor in
Eq.~(\ref{sbk}) suppresses long-distance contributions from the large
$b$ ({\it i.e.}, large $\alpha _{s}(t)$) region, and improves the
applicability of PQCD to $B$ meson decays. We emphasize that the special 
intermediate scales $t\sim O(\sqrt{\bar\Lambda M_B})$ lead to predictions
for penguin-dominated decay modes, such as $B\to\phi K$, which are larger
than those from the factorization and QCD factorization approaches.
When PQCD analyses are extended to $O(\alpha_s^2)$ \cite{CKL}, the hard
scales can be determined more precisely and the scale independence of our
predictions will be improved. The $O(\alpha_s^2)$ corrections to two-body
nonleptonic $B$ meson decays have been computed in the generalized
factorization approach \cite{Ali,CT98}, which indeed improve the scale
independence of the predictions.

For the nonfactorizable amplitudes, the factorization formulas involve the
kinematic variables of all the three mesons \cite{WYL}, and the Sudakov
factor is given by $S=S_{B}S_{\Phi }S_{K}$. Their expressions are 
\begin{equation}
{\cal M}_{e}^{P(q)}={\cal M}_{e3}^{P(q)}+{\cal M}_{e4}^{P(q)}
+{\cal M}_{e5}^{P(q)}+{\cal M}_{e6}^{P(q)}\;,
\label{Me}
\end{equation}
with 
\begin{eqnarray}
{\cal M}_{e4}^{P\left( q\right) } &=&16\pi C_{F}M_{B}^{2}\sqrt{2N_{c}}
\int_{0}^{1}d[x]\int_{0}^{\infty }b_{1}db_{1}b_{2}db_{2}\Phi _{B}
\left(x_{1},b_{1}\right) \Phi _{\phi }\left( x_{2}\right)
\nonumber \\
&&\times \bigg\{ \left[( x_{2}+x_{3})\Phi_{K} (x_{3})
-r_{K} x_{3} \left( \Phi _{K}^{p}(x_{3})
+ \Phi_{K}^{\sigma } ( x_{3} ) \right)
\right] 
\nonumber \\
&&\times E_{e4}^{\left( q\right) \prime }\left( t_{d}^{\left( 1\right)}
\right) h_{d}^{\left( 1\right) }
\left(x_{1},x_{2},x_{3},b_{1},b_{2}\right)
\nonumber \\
&&-\left[ ( 1-x_{2}) \Phi _{K} (x_{3})
-r_{K} x_{3}\left( \Phi _{K}^{p}(x_{3}) -\Phi_{K}^{\sigma}(x_{3})\right)
\right]
\nonumber \\
&& \times E_{e4}^{\left( q\right) \prime }\left( t_{d}^{\left(
2\right) }\right) h_{d}^{\left( 2\right) }\left(
x_{1},x_{2},x_{3},b_{1},b_{2}\right) \bigg\} \;,
\label{Me4} \\
\cr
{\cal M}_{e5}^{P\left( q\right) } &=&16\pi C_{F}M_{B}^{2}\sqrt{2N_{c}}
\int_{0}^{1}d[x]\int_{0}^{\infty }b_{1}db_{1}b_{2}db_{2}\Phi _{B}\left(
x_{1},b_{1}\right) \Phi _{\phi }\left( x_{2}\right)
\nonumber \\
&&\times \bigg\{ \left[ x_{2} \Phi_{K}( x_{3})
- r_{K} x_{3} \left( \Phi _{K}^{p}(x_{3}) -\Phi_{K}^{\sigma}
(x_{3}) \right)  \right] 
\nonumber \\
&&\times E_{e5}^{\left( q\right) \prime }
\left( t_{d}^{\left( 1\right)}\right)
h_{d}^{\left( 1\right) }\left(x_{1},x_{2},x_{3},b_{1},b_{2}\right)
\nonumber \\
&&-\left[ ( 1-x_{2}+x_{3}) \Phi_{K} (x_{3})
-r_{K} x_{3} \left( \Phi _{K}^{p}(x_{3}) +\Phi_{K}^{\sigma}(x_{3})\right)
\right]
\nonumber \\
&& \times E_{e5}^{\left( q\right) \prime }
\left( t_{d}^{\left(2\right) }\right)
h_{d}^{\left( 2\right) }\left(x_{1},x_{2},x_{3},b_{1},b_{2}\right)
\bigg\} \;,
\label{Me5} \\
\cr
{\cal M}_{e6}^{P\left( q\right) } &=&-16\pi C_{F}M_{B}^{2}\sqrt{2N_{c}}
\int_{0}^{1}d[x]\int_{0}^{\infty }b_{1}db_{1}b_{2}db_{2}
\Phi_{B}\left( x_{1},b_{1}\right)   
\nonumber \\
&&\times \bigg\{ \left[r_{\phi} x_{2} 
\left( \Phi_{\phi}^{t}(x_{2}) -\Phi_{\phi}^{s}(x_{2}) \right) 
\Phi_{K}(x_{3}) \right.    
\nonumber \\
&&+r_{K} r_{\phi} \left(x_{2}-x_{3}\right) \left( \Phi_{\phi}^{t}\left(
x_{2}\right) \Phi_{K}^{p}\left( x_{3}\right) +\Phi_{\phi}^{s}\left(
x_{2}\right) \Phi_{K}^{\sigma }\left( x_{3}\right) \right)   
\nonumber \\
&&\left. +r_{K}r_{\phi} ( x_{2}+x_{3}) \left( \Phi_{\phi}^{t}\left(
x_{2}\right) \Phi_{K}^{\sigma }\left( x_{3}\right) -\Phi_{\phi}^{s}\left(
x_{2}\right) \Phi_{K}^{p}\left( x_{3}\right) \right) \right]   
\nonumber \\
&&\times E_{e6}^{\left( q\right) \prime }( t_{d}^{\left( 1\right)
}) h_{d}^{(1)}(x_{1},x_{2},x_{3},b_{1},b_{2})   
\nonumber \\
&&+\left[ r_{\phi}( 1-x_{2})  
\left(\Phi_{\phi}^{t}(x_{2}) +\Phi_{\phi}^{s}(x_{2}) \right)
\Phi_{K}(x_{3}) \right.   
\nonumber \\
&&+r_{K} r_{\phi}( 1-x_{2}-x_{3}) \left( \Phi_{\phi}^{t}\left(
x_{2}\right) \Phi_{K}^{p}\left( x_{3}\right) -\Phi_{\phi}^{s}\left(
x_{2}\right) \Phi_{K}^{\sigma }\left( x_{3}\right) \right)   
\nonumber \\
&&\left. -r_{K} r_{\phi}( 1-x_{2}+x_{3}) \left( \Phi _{\phi}^{t}\left(
x_{2}\right) \Phi_{K}^{\sigma }\left( x_{3}\right) -\Phi_{\phi}^{s}\left(
x_{2}\right) \Phi_{K}^{p}\left( x_{3}\right) \right) \right]   
\nonumber \\
&& \times E_{e6}^{\left( q\right) \prime }\left( t_{d}^{\left(
2\right) }\right) h_{d}^{\left( 2\right) }\left(
x_{1},x_{2},x_{3},b_{1},b_{2}\right) \bigg\} \;,  
\label{Me6}
\end{eqnarray}
where $N_c$ is number of colors and $d[x]$ denotes $dx_1dx_2dx_3$.
The amplitude ${\cal M}_{e3}^{P(q)}$ is the same as ${\cal M}_{e4}^{P(q)}$
but with the Wilson coefficient $a_{3}^{(q)\prime}$. The nonfactorizable
annihilation amplitudes are given by 
\begin{equation}
{\cal M}_{a}^{P(q)}={\cal M}_{a3}^{P(q)}+{\cal M}_{a5}^{P(q)}\;,
\label{Ma}
\end{equation}
with 
\begin{eqnarray}
{\cal M}_{a3}^{P\left( q\right) } &=&-16\pi C_{F}M_{B}^{2}\sqrt{2N_{c}}
\int_{0}^{1}d[x]\int_{0}^{\infty }b_{1}db_{1}b_{2}db_{2}\Phi _{B}\left(
x_{1},b_{1}\right)   
\nonumber \\
&&\times \bigg\{ \left[ \left( 1-x_{3}\right) \Phi _{\phi }\left(
x_{2}\right) \Phi _{K}\left( x_{3}\right) \right.    
\nonumber \\
&&+r_{K}r_{\phi}\left(1-x_{2}-x_{3}\right) 
\left( \Phi_{\phi}^{t}\left(x_{2}\right) \Phi_{K}^{p}\left( x_{3}\right) 
-\Phi_{\phi}^{s}\left(x_{2}\right) \Phi_{K}^{\sigma }\left( x_{3}\right) 
\right)   
\nonumber \\
&& \left. -r_{K}r_{\phi} ( 1+x_{2}-x_{3}) 
\left( \Phi_{\phi}^{t}( x_{2}) \Phi _{K}^{\sigma }(x_{3}) 
-\Phi_{\phi}^{s}( x_{2}) \Phi_{K}^{p}( x_{3}) \right) \right]
\nonumber \\
&&\times E_{a3}^{\left( q\right) \prime}\left( t_{f}^{\left( 1\right)
}\right) h_{f}^{\left( 1\right) }\left(
x_{1},x_{2},x_{3},b_{1},b_{2}\right)   
\nonumber \\
&&-\left[ x_{2} \Phi_{\phi}( x_{2}) \Phi_{K}(x_{3}) \right.\nonumber \\
&&-r_{K}r_{\phi}(1-x_{2}-x_{3}) 
\left( \Phi_{\phi}^{t}(x_{2}) \Phi_{K}^{p} (x_{3}) + \Phi_{\phi}^{s}
(x_{2}) \Phi_{K}^{\sigma }(x_{3}) \right)   
\nonumber \\
&&\left. + r_{K}r_{\phi} 
(1-x_2+x_3)\left( \Phi_{\phi}^{t}(x_{2}) \Phi_{K}^{\sigma }( x_{3} ) 
+ ( 3+x_{2}-x_{3})  
\Phi_{\phi}^{s} (x_{2}) \Phi_{K}^{p} ( x_{3}) \right) \right]  
\nonumber \\
&& \times E_{a3}^{\left( q\right) \prime }\left( t_{f}^{\left(
2\right) }\right) h_{f}^{\left( 2\right) }\left(
x_{1},x_{2},x_{3},b_{1},b_{2}\right) \bigg\} ,  
\label{Ma4} \\
\cr
{\cal M}_{a5}^{P\left( q\right) } &=&-16\pi C_{F}M_{B}^{2}\sqrt{2N_{c}}%
\int_{0}^{1}d[x]\int_{0}^{\infty }b_{1}db_{1}b_{2}db_{2}\Phi _{B}\left(
x_{1},b_{1}\right)  
\nonumber \\
&&\times \bigg\{ \left[ r_{\phi}x_{2} 
 \left( \Phi _{\phi}^{t}\left( x_{2}\right) +\Phi _{\phi}^{s}\left(
x_{2}\right) \right) \Phi_{K}(x_3)\right.   
\nonumber \\
&&\left. - r_{K} ( 1-x_{3}) \Phi_{\phi}(x_{2}) \left( \Phi_{K}^{p}(x_{3}) 
-\Phi _{K}^{\sigma }(x_{3}) \right)
\right] 
\nonumber\\
& &\times E_{a5}^{\left( q\right) \prime }\left( t_{f}^{\left( 1\right)
}\right) h_{f}^{\left( 1\right) }\left(
x_{1},x_{2},x_{3},b_{1},b_{2}\right)  
\nonumber \\
&&+\left[ r_{\phi} ( 2-x_{2}) 
\left( \Phi_{\phi}^{t}(x_{2}) +\Phi_{\phi}^{s}(x_{2})
\right) \Phi_{K}(x_3) \right.  
\nonumber \\
&&\left. - r_{K}(1+x_{3}) \Phi_{\phi}(x_{2}) \left(
\Phi_{K}^{p}(x_{3}) -\Phi_{K}^{\sigma}(x_{3}) \right) \right]  
\nonumber \\
&& \times E_{a5}^{\left( q\right) \prime }\left( t_{f}^{\left(
2\right) }\right) h_{f}^{\left( 2\right) }\left(
x_{1},x_{2},x_{3},b_{1},b_{2}\right) \bigg\} \;.  
\label{Ma6}
\end{eqnarray}
The amplitude ${\cal M}_{a}$ is the same as ${\cal M}_{a3}^{P(u)}$ but
with Wilson coefficient $a_{1}^{\prime}$. 

The evolution factors are given by 
\begin{eqnarray}
E_{ei}^{\left( q\right) \prime }\left( t\right) &=&\alpha _{s}\left(
t\right) a_{i}^{\left( q\right)\prime}(t)S\left( t\right)|_{b_{3}=b_{1}}\;,
 \nonumber \\
E_{ai}^{\left( q\right) \prime }\left( t\right)
&=&\alpha _{s}\left( t\right)
a_{i}^{\left( q\right)\prime}(t)S\left( t\right)|_{b_{3}=b_{2}}\;.
\label{Eeaprim}
\end{eqnarray}
The hard functions $h^{(j)}$, $j=1$ and 2, are written as 
\begin{eqnarray}
h_{d}^{(j)} &=&\left[ \theta (b_{1}-b_{2})K_{0}\left( DM_{B}b_{1}\right)
I_{0}\left( DM_{B}b_{2}\right) \right.  
\nonumber \\
&&\quad \left. +\theta (b_{2}-b_{1})K_{0}\left( DM_{B}b_{2}\right)
I_{0}\left( DM_{B}b_{1}\right) \right]  \nonumber \\
&&\times K_{0}(D_{j}M_{B}b_{2})\;,\;\;\;\;\;\;\;\;\;\;\;\;\;\;\;\;\;\;\;
\mbox{for $D^2_{j} \geq 0$}  
\nonumber \\
&&\times \frac{i\pi }{2}H_{0}^{(1)} 
\left(\sqrt{|D_{j}^{2}|}M_{B}b_{2}\right)\;,\;\;\;\; 
\mbox{for $D^2_{j} \leq 0$}\;,  
\label{hjd} \\
h_{f}^{(j)} &=&\frac{i\pi }{2}\left[ \theta (b_{1}-b_{2})H_{0}^{(1)}\left(
FM_{B}b_{1}\right) J_{0}\left( FM_{B}b_{2}\right) \right.  
\nonumber \\
&&\quad \left. +\theta (b_{2}-b_{1})H_{0}^{(1)}\left( FM_{B}b_{2}\right)
J_{0}\left( FM_{B}b_{1}\right) \right]  
\nonumber \\
&&\times K_{0}(F_{j}M_{B}b_{1})\;,\;\;\;\;\;\;\;\;\;\;\;\;\;\;\;\;\;\;\;
\mbox{for $F^2_{j} \geq 0$}  
\nonumber \\
&&\times \frac{i\pi }{2}H_{0}^{(1)}
\left(\sqrt{|F_{j}^{2}|}M_{B}b_{1}\right)\;,\;\;\;\; 
\mbox{for $F^2_{j} \leq 0$}\;,  
\label{hjf}
\end{eqnarray}
with the variables 
\begin{eqnarray}
D^{2} &=& x_{1}x_{3}\;,  
\nonumber \\
D_{1}^{2} &=&(x_{1}-x_{2})x_{3}\;,  
\nonumber \\
D_{2}^{2} &=&-(1-x_{1}-x_{2})x_{3}\;, \\
F^{2} &=& x_{2} \left( 1-x_{3}\right) \;,  
\nonumber \\
F_{1}^{2} &=&(x_{1}-x_{2}) \left( 1-x_{3}\right)\;,  
\nonumber \\
F_{2}^{2} &=&x_{1}+x_{2}+\left(1- x_{1}-x_{2}\right) 
\left( 1-x_{3}\right)\;.
\label{DF}
\end{eqnarray}
The hard scales $t^{(j)}$ are chosen as 
\begin{eqnarray}
t_{d}^{(1)} &=&{\rm max}\left(DM_{B},\sqrt{|D_{1}^{2}|}M_{B},1/b_{1},
1/b_{2}\right)\;,
\nonumber \\
t_{d}^{(2)} &=&{\rm max}\left(DM_{B},\sqrt{|D_{2}^{2}|}M_{B},1/b_{1},
1/b_{2}\right)\;,
\nonumber \\
t_{f}^{(1)} &=&{\rm max}\left(FM_{B},\sqrt{|F_{1}^{2}|}M_{B},1/b_{1},
1/b_{2}\right)\;,
\nonumber \\
t_{f}^{(2)} &=&{\rm max}\left(FM_{B},\sqrt{|F_{2}^{2}|}M_{B},1/b_{1},
1/b_{2}\right)\;.
\end{eqnarray}
In the above factorization formulas the Wilson coefficients are defined by 
\begin{eqnarray*}
a_{1} &=&C_{1}+\frac{C_{2}}{N_{c}}\;, \;\;\;\;
a_{1}^{\prime} =\frac{C_{1}}{N_{c}}\;, \\
a_{2} &=&C_{2}+\frac{C_{1}}{N_{c}}\;, \;\;\;\;
a_{2}^{\prime} =\frac{C_{2}}{N_{c}}\;, \\
a_{3}^{(q)} &=&C_{3}+\frac{C_{4}}{N_{c}}+\frac{3}{2}e_{q}\left( C_{9} 
+\frac{C_{10}}{N_{c}}\right) \;, \\
a_{3}^{(q)\prime} &=&\frac{1}{N_{c}} \left( C_{3}+\frac{3}{2}
e_{q}C_{9}\right) \;, \\
a_{4}^{(q)} &=&C_{4}+\frac{C_{3}}{N_{c}}+\frac{3}{2}e_{q} \left( C_{10}+
\frac{C_{9}}{N_{c}}\right) \;, \\
a_{4}^{(q)\prime } &=&\frac{1}{N_{c}} \left( C_{4}+\frac{3}{2}
e_{q}C_{10}\right);, \\
a_{5}^{(q)} &=&C_{5}+\frac{C_{6}}{N_{c}}+\frac{3}{2}e_{q} \left( C_{7}
+\frac{C_{8}}{N_{c}}\right) \;, \\
a_{5}^{(q)\prime} &=&\frac{1}{N_{c}}\left( C_{5}+\frac{3}{2}e_{q}C_{7}
\right)\;, \\
a_{6}^{(q)} &=&C_{6}+\frac{C_{5}}{N_{c}}+\frac{3}{2}e_{q} \left( C_{8}
+\frac{C_{7}}{N_{c}}\right) \;, \\
a_{6}^{(q)\prime} &=&\frac{1}{N_{c}}\left( C_{6}+\frac{3}{2}e_{q}C_{8}
\right)\;, \\
a_{e}^{(q)} &=&a_{3}^{(q)}+a_{4}^{(q)}+a_{5}^{(q)}\;.
\end{eqnarray*}

\section{Numerical Analysis}

For the $B$ meson distribution amplitude, we adopt the model 
\cite{{Keum:2001ph},{Keum:2001wi}} 
\begin{eqnarray}
\Phi_{B}(x,b)=N_{B}x^{2}(1-x)^{2}\exp \left[ -\frac{1}{2}
\left( \frac{xM_{B}}{\omega _{B}}\right) ^{2}
-\frac{\omega _{B}^{2}b^{2}}{2}\right]
\label{bw} \;,
\end{eqnarray}
with the shape parameter $\omega_{B}=0.4$ GeV. The normalization
constant $N_{B}= 91.784$ GeV is related to the decay constant $f_{B}=190$
MeV (in the convention $f_{\pi}=130$ MeV). As stated before, $\Phi_B$ has
a sharp peak at $x\sim \bar\Lambda/M_B\sim 0.1$. The $\phi$ and $K$ meson
distribution amplitudes are derived from QCD sum rules \cite{PB2,PB1},
\begin{eqnarray}
\Phi _{\phi }\left( x\right)  &=&\frac{3f_\phi}{\sqrt{2N_{c}}}x(1-x)\;,
\label{phi2}\\
\Phi _{\phi}^{t}\left( x\right)  &=&\frac{f_\phi^T}{2\sqrt{2N_{c}}}
\bigg\{ 3(1-2x)^{2}+0.21\left[3-30(1-2x)^{2}+35(1-2x)^{4}\right]
\nonumber\\
&&+0.69\left( 1+(1-2x)\ln \frac{x}{1-x}\right) \bigg\} \;,
\label{phi3t}\\
\Phi _{\phi}^{s}\left( x\right)  &=&\frac{f_\phi^T}{4\sqrt{2N_{c}}}
\left[ 3(1-2x)(4.5-11.2x+11.2x^{2})+1.38\ln \frac{x}{1-x}\right] \;, 
\label{phi3s}\\
\Phi _{K}(x) &=&\frac{3f_K}{\sqrt{2N_{c}}}
x(1-x)\left\{1+0.51(1-2x)+0.3[5(1-2x)^{2}-1]\right\}\;, 
\label{kk}\\
\Phi _{K}^{p}(x) &=&\frac{f_K}{2\sqrt{2N_{c}}}\left[
1+0.24C_{2}^{1/2}(1-2x)-0.11C_{4}^{1/2}(1-2x)\right] \;, 
\label{kp}\\
\Phi _{K}^{\sigma }(x) &=&\frac{f_K}{2\sqrt{2N_{c}}}(1-2x)\left[
1+0.35(10x^{2}-10x+1)\right] \;,
\label{ks}
\end{eqnarray}
with the Gegenbauer polynomials
\begin{eqnarray}
C_2^{1/2}(\xi)=\frac{1}{2}[3\xi^2-1]\;,\;\;\;
C_4^{1/2}(\xi)=\frac{1}{8}[35 \xi^4 -30 \xi^2 +3]\;.
\end{eqnarray}
To derive the coefficients of the Gegenbauer polynomials, we have assumed
$M_K=0.49$ GeV and $m_0=1.7$ GeV. The terms $1-2x$, rendering the kaon 
distribution amplitudes a bit asymmetric, corresponds to the $SU(3)$
symmetry breaking effect.
We employ $G_{F}=1.16639\times 10^{-5}$ GeV$^{-2}$, the Wolfenstein
parameters $\lambda =0.2196$, $A=0.819$, and $R_{b}=0.38$, the unitarity
angle $\phi_{3}=90^{o}$, the masses $M_{B}=5.28$ GeV and  
$M_{\Phi}=1.02$ GeV, the decay constants $f_{\phi}=237$ MeV, 
$f_{\phi}^{T}=220$ MeV and $f_{K}=160$ MeV, and the 
${\bar{B}}_{d}^{0}$ ($B^{-}$) meson lifetime $\tau_{B^{0}}=1.55$ ps 
($\tau_{B^{-}}=1.65$ ps) \cite{PDG}. Note that the $B\to\phi K$
branching ratios are insensitive to the variation of $\phi_3$.

We present values of the factorizable and nonfactorizable amplitudes from
the emission and annihilation topologies in Table III. Contributions from
twist-2 and two-parton twist-3 distribution amplitudes are displayed 
separately. It is found that the latter, not power-suppressed, are in 
fact more important for $f_\phi F_e^P$. According to the power counting
in Sec.~III, the twist-2 contributions to the annihilation amplitudes
$f_B F^P_a$ are negligible. This has been explicitly confirmed in Table
III. As expected, the factorizable amplitudes $f_\phi F_e^P$ dominate,
and the annihilation amplitudes $f_B F_a^P$ are almost imaginary and
their magnitudes are only few times smaller than $f_\phi F_e^P$.
The nonfactorizable amplitudes $M_e^P$ and $M_a^P$ are down by a power
of $\bar\Lambda/M_B\sim 0.1$ compared to the factorizable ones
$f_\phi F_e^P$ and $f_B F_a^P$, respectively. The cancellation between
the twist-2 and twist-3 contributions makes them even smaller. $M_a$ and
$f_B F_a$ from the operators $O_{1,2}$ are of the same order because of
the partial cancellation between the two terms in the factorization
formula for $F_a$ (helicity suppression).


We demonstrate the importance of penguin enhancement in Table IV. 
It has been known that the RG evolution of the Wilson coefficients 
$C_{4,6}(t)$ dramatically increases as $t<m_b/2$, while that of 
$C_{1,2}(t)$ almost remains constant \cite{REVIEW}. With this penguin 
enhancement of about 40\%, the branching ratios of the $B\to K\pi$ 
decays, dominated by penguin contributions, are about four times 
larger than those of the $B\to\pi\pi$ decays, which are dominated by 
tree contributions. This is the reason we can explain the observed 
$B\to K\pi$ and $\pi\pi$ branching ratios using a smaller unitarity 
angle $\phi_3<90^o$ \cite{Keum:2001ph,Keum:2001wi}. In the factorization 
approach \cite{BSW} and in the QCD factorization approach \cite{BBNS}, 
it is assumed that factorizable contributions are not calculable. The 
leading contribution to a nonleptonic decay amplitude is then expressed
as a convolution of a hard part with a form factor and a meson
distribution amplitude. In both approaches the hard scale is $m_b$ and the 
intermediate scale $\bar\Lambda M_B$ can not appear, so that the 
dynamical enhancement of penguin contributions does not exist. 
To accommodate the $B\to K\pi$ data in the factorization and QCD 
factorization approaches, one relies on the chiral enhancement
by increasing the mass $m_0$ to a large value $m_0\sim 3$ GeV, or
on a large unitarity angle $\phi_3\sim 120^o$ \cite{WS}, which leads to
constructive (destructive) interference between penguin and emission
amplitudes for the $B\to K\pi$ ($B\to\pi\pi$) decays.

Whether dynamical enhancement or chiral enhancement is essential for 
the penguin-dominated decay modes can be tested by measuring the 
$B\to \phi K$ modes. In these modes penguin contributions dominate, 
and their branching ratios are almost independent of the 
angle $\phi_3$. Since $\phi$ is a vector meson, the mass $m_0$ is 
replaced by the $\phi$ meson mass $M_\phi\sim 1$ GeV, and chiral 
enhancement does not exist. Annihilation contributions can not
enhance the $B\to \phi K$ branching ratios too much, because they 
are assumed to be a $1/m_b$ effect in the QCD factorization 
approach \cite{BBNS}. In the PQCD approach annihilation amplitudes 
reach 40\%, which is reasonable according to our power counting. 
However, they, being mainly imaginary, are not responsible for the large
$B\to\phi K$ branching ratios as shown in Table IV. If the 
$B\to\phi K$ branching ratios 
are around $4\times 10^{-6}$ \cite{HMW,CY},
the chiral enhancement may be essential. If the branching ratios are 
around $10 \times 10^{-6}$, the dynamical 
enhancement may be essential. Therefore, the $B\to\phi K$ decays are 
the appropriate modes to distinguish the QCD and PQCD factorization 
approaches. The branching ratios of $B_d^0\to \phi K^0$ and of
$B^\pm\to \phi K^\pm$ are almost euqal. We have also evaluated the CP
asymmetries of the $B\to \phi K$ decays, and found that they are not
significant: their maxima, appearing at $\phi_3\sim 90^o$, are less
than 2\%.

We emphasize that $m_0(\mu)$, appearing along with the twist-3 kaon
distribution amplitudes, is defined at the factorization scale $1/b$ as
low as 1 GeV in the PQCD formalism \cite{CL,YL}. Hence, its value
should be located within $1.6\pm 0.2$ GeV \cite{PB2}. Between the hard
scale and the factorization scale, there is the Sudakov evolution.
In the QCD factorization approach, $m_0(\mu)$ defined at $m_b$
is as large as 3 GeV, leading to chiral enhancement. It has
been argued that $m_0(\mu)$ and $a_6(\mu)$ form a scale-independent
product (that is, $m_0(\mu)$ increases, while $a_6(\mu)$ decreases with
$\mu$), such that chiral and dynamical enhancements can not be
distinguished \cite{CY2}. However, dynamical enhancement exists in both
twist-2 and twist-3 contributions, but chiral enhancement exists only in
twist-3 ones (see Eq.~(\ref{Fe})). Therefore, they are indeed different
mechanism.

At last, we examine the uncertainty from the variation of the hard 
scales $t$, which provides the informaiton of higher-order corrections 
to the hard amplitudes. We notice that this is the major source of 
the theoretical uncertainty. The values of $\omega_B$ and $m_0$ have been
fixed at around 0.4 GeV and 1.7 GeV, respectively, which are preferred by
the $B\to K\pi$, $\pi\pi$ data. The light meson distribution amplitudes
have been fixed more or less in QCD sum rules. The possible 30\%
variation of the coefficients of the Gegenbauer polynomials lead to minor
changes of our predictions. Since the analyses of the $B\to\pi$ and
$B\to K$ form factors are the same, we constraint the ranges of
the hard scales $t$, such that our predictions for the $B\to K\pi$
branching ratios are within the data uncertainties. The resultant
approximate range of the hard scales $t_e$ is given by
\begin{eqnarray}
{\rm max}(0.75\sqrt{x_{3}}M_{B},1/b_{1},1/b_{3})<t_{e}^{(1)}<
{\rm max}(1.25\sqrt{x_{3}}M_{B},1/b_{1},1/b_{3})\;,
\nonumber \\
{\rm max}(0.75\sqrt{x_{1}}M_{B},1/b_{1},1/b_{3})<
t_{e}^{(2)} <{\rm max}(1.25\sqrt{x_{1}}M_{B},1/b_{1},1/b_{3})\;.
\end{eqnarray}
Note that the coefficients of the factorization scales $1/b$, associated
with the definition of the meson distribution amplitudes, do not
change. The variation of the other hard scales $t$ is similar, but
does not affect the results very much.
The theoretical uncertainty for the $B\to\phi K$ branching ratios
in Eq.~(\ref{phikb}) is then obtained.

\section{CONCLUSION}

In this paper we have shown that a leading-power PQCD formalism should 
contain contributions from both twist-2 and two-parton twist-3 
distribution amplitudes. Threshold and $k_T$ resummations are essential 
for infrared finite PQCD analyses of $B$ meson decays. Without 
Sudakov suppression from these resummations, all topologies of decay 
amplitudes possess infrared (logarithmic or linear) divergences. We 
have explained the power counting rules for the factorizable (also
nonfactorizable) emission and annihilation amplitudes under the
sufficiently strong Sudakov effects. The annihilation and nonfactorizable
amplitudes are suppressed by $2m_0/M_B$ and by $\bar\Lambda/M_B$ in the
heavy quark limit, respectively, compared to the factorizable emission
ones. For the physical mass $M_B\sim 5$ GeV, the former should
be taken into account. In the PQCD formalism the annihilation amplitudes
can be calculated in the same way as the emission ones without
introducing any new free parameters. Hence, our formalism has more
precise control on the annihilation effects than the QCD factorization
approach. Annihilation contributions of 40\% at the
amplitude level are reasonable according to our power counting. 
However, these amplitudes are not responsible for the large 
$B\to\phi K$ branching ratios, since they are mainly imaginary.

We have emphasized that exclusive heavy meson decays are characterized 
by a lower scale $\bar\Lambda M_B$, for which penguin contributions 
are dynamically enhanced. This enhancement renders penguin-dominated 
decay modes acquire branching ratios larger than those from the 
factorization and QCD factorization approaches, even when the
final-state particle is a vector meson. We have proposed the
$B\to\phi K$ decays as the ideal modes to test the importance of this
mechanism. If their branching ratios are as large as
$10 \times 10^{-6}$ (independent of the unitarity angle $\phi_3$),
dynamical enhancement will gain a convincing support. The answer will
become clear, when the consistency among the BaBar,
Belle, and CLEO data is achieved. We have also found that the
CP asymmetries in the $B\to\phi K$ modes are vanishingly small
(less than 2\%).

\vskip 1.0cm

We thank S. Brodsky, H.Y. Cheng, L. Dixon, G. Hiller, H. Quinn, A.I.
Sanda and K.C. Yang for helpful discussions. The work was supported in
part by Grant-in Aid for Special Project Research (Physics of CP
Violation), by Grant-in Aid for Scientific Research from the Ministry of
Education, Science and Culture of Japan. The work of H.N.L. was supported
in part by the National Science Council of R.O.C. under the Grant No.
NSC-89-2112-M-006-033 and by Theory Group of SLAC. The work of Y.Y.K was
supported in part by Natioanl Science Council of R.O.C. under the Grant
No. NSC-90-2811-M-002.

\appendix

\section{TWO-PARTON DISTRIBUTION AMPLITUDES}

\subsection{$B$ Meson Distribution Amplitudes}

The $B$ meson distribution amplitudes are written as \cite{TLS,GN}
\begin{equation}
\frac{(\not{P}_1+M_B)\gamma_5}{\sqrt{2N_c}}\Phi_B(x,b)\;,\;\;\;\;
\frac{(\not{P}_1+M_B)\gamma_5}{\sqrt{2N_c}}
\frac{\not n_+-\not n_-}{\sqrt{2}}\bar\Phi_B(x,b)\;.
\end{equation}
with the dimensioless vectors $n_+=(1,0,0_{T})$ and $n_-=(0,1,0_{T})$.
As shown in \cite{TLS}, the contribution from $\bar\Phi_B$ is negligible,
after taking into account the equation of motion between $\Phi_B$ and 
$\bar\Phi_B$. Hence, we consider only a single $B$ meson distribution 
amplitude in the heavy quark limit in this work. As the transverse
extent $b$ approaches zero, the $B$ meson distribution amplitude
$\Phi_{B}(x,b)$ reduces to the standard parton model
$\Phi_{B}(x)=\Phi_{B}(x,b=0)$, which satisfies the normalization
\begin{equation}
\int_{0}^{1}\Phi_{B}(x)dx=\frac{f_{B}}{2\sqrt{2N_{c}}}\;.
\label{dco}
\end{equation}

\subsection{$\phi$ Meson Distribution Amplitudes}

To define the $\phi$ meson distribution amplitudes, we consider the
following nonlocal matrix elements \cite{PB1}, 
\begin{eqnarray}
\langle 0|{\bar s}(0)\gamma_\mu s(z)|\phi(P_2)\rangle&=&
f_{\phi}M_{\phi}\left[P_{2\mu}\frac{\epsilon\cdot z}{P_2\cdot z} \int_0^1
dx_2 e^{-ix_2P_2\cdot z}\phi_{\parallel}(x_2) \right.
\nonumber\\
& &+\epsilon_{T\mu}\int_0^1 dx_2e^{-ix_2P_2\cdot z}g^{(v)}_T(x_2)
\nonumber \\
& &\left. -\frac{1}{2}z_\mu\frac{\epsilon\cdot z}{(P_2\cdot z)^2}
M^2_{\phi}\int_0^1 dx_2 e^{-ix_2P_2\cdot z}g_3(x_2)\right]\;,
\label{v} \\
\langle 0|{\bar s}(0)\sigma_{\mu\nu}s(z)|\phi(P_2)\rangle&=&
if^T_{\phi}\left[(\epsilon_{T\mu}P_{2\nu}-\epsilon_{T\nu}P_{2\mu})
\int_0^1dx_2 e^{-ix_2P_2\cdot z}\phi_T(x_2)\right.
\nonumber \\
& &+(P_{2\mu} z_\nu-P_{2\nu} z_\mu)\frac{\epsilon\cdot z}{(P_2\cdot z)^2}
M^2_{\phi}\int_0^1 dx_2 e^{-ix_2P_2\cdot z}h^{(t)}_{\parallel}(x_2)
\nonumber \\
& &\left. +\frac{1}{2}(\epsilon_{T\mu}z_\nu-\epsilon_{T\nu}z_{\mu})
\frac{M^2_{\phi}}{P_2\cdot z}\int_0^1 dx_2 e^{-ix_2P_2\cdot z}h_3(x_2)
\right]\;,
\label{t} \\
\langle 0|{\bar s}(0)I s(z)|\phi(P_2)\rangle&=& \frac{i}{2}
\left(f^T_{\phi}-f_{\phi}\frac{2m_s}{M_{\phi}}\right) \epsilon\cdot
zM^2_{\phi}\int_0^1 dx_2 e^{-ix_2P_2\cdot z} h^{(s)}_{\parallel}(x_2)\;,
\label{s}
\end{eqnarray}
where $f_{\phi}$ and $f^T_{\phi}$ are the decay constants of the $\phi$
meson with longitudinal and transverse polarizations, respectively, 
$\epsilon_T$ the transverse polarization vector, $x_2$ the momentum 
associated with the $s$ quark at the coordinate $z\propto (0,1,0_T)$, 
and $m_s$ the $s$ quark mass. The explicit expressions of the distribution 
amplitudes $\phi$, $g$ and $h$ with unity normalization are referred to 
\cite{PB1}.

The contributions from the distribution amplitudes $g_T^{(v)}$, $\phi_T$
and $h_3$ vanish for two-body $B$ meson decays, in which only
longitudinally polarized $\phi$ mesons are produced. The contributions
from $\phi_{\parallel}$, $h^{(t)}_{\parallel}$, $h^{(s)}_{\parallel}$
and $g_3$ are twist-2, twist-3, twist-3, and twist-4, respectively.
It is easy to confirm that $g_3$ does not contribute to the factorizable
emission and annihilation amplitudes. The term proportional to the small
ratio $2m_{s}/M_{\phi}$ in Eq.~(\ref{s}) is negligible. Therefore, up to
twist 3, we consider the following three $\phi$ meson final-state
distribution amplitudes,
\begin{eqnarray}
\frac{M_{\phi}\not{\epsilon}}{\sqrt{2N_c}}\Phi_{\phi}(x_{2})\;,\;\;\;
\frac{\not{\epsilon}\not{P}_{2}}{\sqrt{2N_c}}\Phi_{\phi}^{t}(x_{2})\;,\;\;\;
\frac{M_{\phi}}{\sqrt{2N_c}}\Phi_{\phi}^{s}(x_{2})\;, 
\label{sph}
\end{eqnarray}
with 
\begin{eqnarray}
\Phi_{\phi}=\frac{f_{\phi}}{2\sqrt{2N_{c}}}\phi_{\parallel}\;,\;\;\;
\Phi_{\phi}^{t}=\frac{f_{\phi}^T}{2\sqrt{2N_c}}h_{\parallel}^{(t)}\;,\;\;\;
\Phi_{\phi}^{s}=\frac{f_{\phi}^T}{4\sqrt{2N_{c}}}\frac{d}{dx}
h_{\parallel}^{(s)}\;.
\end{eqnarray}
The spin structures associated with the $\phi$ meson distribution
amplitudes can be derived from Eqs.~(\ref{v})-(\ref{s}).

\subsection{Kaon Distribution Amplitudes}

The general expressions of the relevant nonlocal matrix elements
for a kaon are given by \cite{PB2}, 
\begin{eqnarray}
\langle 0|{\bar s}(0)\gamma_5\gamma_\mu u(z)|K(P_3)\rangle&=&
-if_KP_{3\mu}\int_0^1 dx_3 e^{-ix_3P_3\cdot z}\phi_v(x_3)
\nonumber \\
& &-\frac{i}{2}f_KM_K^2\frac{z_\mu}{P_3\cdot z} \int_0^1 dx_3
e^{-ix_3P_3\cdot z}g_K(x_3)\;,
\label{pv} \\
\langle 0|{\bar s}(0)\gamma_5 u(z)|K(P_3)\rangle&=& -if_Km_{0}
\int_0^1 dx_3e^{-ix_3P_3\cdot z}\phi_p(x_3)\;,
\label{ps} \\
\langle 0|{\bar s}(0)\gamma_5\sigma_{\mu\nu}u(z)|K(P_3)\rangle&=&
\frac{i}{6}f_Km_{0}\left(1-\frac{M_K^2}{m_{0}^2}\right)
(P_{3\mu}z_\nu-P_{3\nu}z_\mu)
\nonumber\\
& & \times\int_0^1 dx_3 e^{-ix_3P_3\cdot z}\phi_\sigma(x_3)\;,
\label{pt}
\end{eqnarray}
with the mass 
\begin{equation}
m_{0}=\frac{M_K^2}{m_s+m_d}\;.
\end{equation}
$f_K$ is the kaon decay constant and $x_3$ the momentum fraction 
associated with the $u$ quark at the coordinate $z\propto (1,0,0_T)$. 
The explicit expression of the wave fucntions $\phi$ and $g_K$ with 
unit normalization are referred to \cite{PB2}.

The contributions from the distribution amplitudes $\phi_v$, $\phi_p$,
$\phi_\sigma$ and $g_K$ are twist-2, twist-3, twist-3, and twist-4,
respectively. Note that $g_K$ does not contribute to factorizable emission
and annihilation amplitudes. Hence, the factorizable annihilation
amplitude in Eq.~(\ref{Fa4}) is complete in the $r^2$ terms. We consider
the following three kaon final-state distribution amplitudes 
\begin{eqnarray}
\frac{\gamma_{5}\not{P}_{3}}{\sqrt{2N_{c}}}\Phi_{K}\;,\;\;\;
\frac{m_{0}\gamma _{5}}{\sqrt{2N_{c}}}\Phi_{K}^{p}\;,\;\;\;
\frac{m_{0}\gamma _{5}(\not n_- \not n_+-1)}{\sqrt{2N_{c}}}
\Phi_{K}^{\sigma }\;,
\end{eqnarray}
with 
\begin{eqnarray}
\Phi_{K}(x)=\frac{f_{K}}{2\sqrt{2N_{c}}}\phi_{v}(x)\;,\;\;\;
\Phi_{K}^{p}(x)=\frac{f_{K}}{2\sqrt{2N_{c}}}\phi_{p}(x)\;,\;\;\;
\Phi_{K}^{\sigma}(x)=\frac{f_{K}}{12\sqrt{2N_{c}}}\frac{d}{dx}
\phi_{\sigma }(x)\;,
\end{eqnarray}
where the term $M_K^2/m_0^2$ in Eq.~(\ref{pt}) has been neglected. The
spin structures associated with the kaon distribution amplitudes can be
derived from Eqs.~(\ref{pv})-(\ref{pt}).


\vskip 1.0cm

{\bf \Large Figure Captions}
\vspace{10mm}

\noindent
{\bf FIG. 1}:
Leading contribution to the $B \to \phi K $ decays, where
$H$ denotes the hard amplitude and $k_i$, $i=1$, 2, 3
are the parton momenta.
\vskip 0.5cm

\noindent
{\bf FIG. 2}:
Lowest-order diagrams for the $B \to \pi, K$ transition form factors.

\newpage
\begin{table}[ht]
\vspace*{0.5cm}
\caption{$B \to \pi,K$ transition form factors without threshold
resummation (column A) and with threshold resummation (column B).
}\label{table-1} \vspace{10mm}
\begin{center}
\begin{tabular}{|c||c|c||c|c||}
\hline
Form factors & \multicolumn{2}{c||}{$F^{B\pi }(0)$} & \multicolumn{2}{c||}
{$F^{BK}(0)$} \\ \hline\hline ~~~$\omega _{B}$ (GeV) 
& A~ & B~ & A~ & B~ \\ \hline\hline 
0.35 & 0.603~~ & 0.356~~ & 0.703~~ & 0.430~~ \\ \hline
0.36 & 0.579~~ & 0.343~~ & 0.675~~ & 0.413~~ \\ \hline 
0.37 & 0.557~~ & 0.330~~ & 0.647~~ & 0.398~~ \\ \hline 
0.38 & 0.535~~ & 0.318~~ & 0.621~~ & 0.382~~ \\ \hline 
0.39 & 0.515~~ & 0.306~~ & 0.597~~ & 0.368~~ \\ \hline
0.40 & 0.496~~ & 0.295~~ & 0.574~~ & 0.354~~ \\ \hline 
0.41 & 0.478~~ & 0.285~~ & 0.552~~ & 0.342~~ \\ \hline 
0.42 & 0.461~~ & 0.275~~ & 0.532~~ & 0.329~~ \\ \hline 
0.43 & 0.445~~ & 0.266~~ & 0.512~~ & 0.318~~ \\ \hline
0.44 & 0.429~~ & 0.257~~ & 0.494~~ & 0.307~~ \\ \hline 
0.45 & 0.414~~ & 0.248~~ & 0.476~~ & 0.296~~ \\ \hline
\end{tabular}
\end{center}
\end{table}

\vskip 1.0cm
\begin{table}[ht]
\vspace*{0.5cm}
\caption{{ Time-like form factors $F^{\phi K}$}
\label{table-2}}
\vspace{10mm}
\begin{center}
\begin{tabular}{|c||c||c||c|c||c|c||}
\hline
Form factors & $F^{B\pi }(0)$ & $F^{BK}(0)$ & \multicolumn{2}{c||}
{$F_{(V-A)}^{\phi K}$ $\times 10^{2}$} &
\multicolumn{2}{c||}{$F_{(V+A)}^{\phi K}$ $\times 10^{2}$~} \\
\hline\hline ~~~$m_{0}$(GeV) & Re & Re & Re~~~~ & Im~~ & Re~~~~ & Im~~ \\
\hline\hline 1.4 & 0.295 & 0.312 & 1.49 & 0.56 & -2.58 & 12.72 \\
\hline 1.5 & 0.308 & 0.326 & 1.61 & 0.57 & -3.06 & 13.13~ \\
\hline 1.6 & 0.320 & 0.340 & 1.71 & 0.62 & -3.20 & 13.41 \\ \hline
1.7 & 0.333 & 0.354 & 1.78 & 0.63 & -3.48 & 13.52 \\ \hline 1.8 &
0.345 & 0.368 & 1.92 & 0.68 & -3.79 & 13.88 \\ \hline 1.9 & 0.357
& 0.383 & 2.02 & 0.73 & -3.99 & 14.67 \\ \hline 2.0 & 0.370 &
0.396 & 2.10 & 0.74 & -4.40 & 14.82 \\ \hline
\end{tabular}
\end{center}
\end{table}

\newpage
\begin{table}[ht]
\vspace*{0.5cm}
\caption{{  Twist-2 and higher-twist contributions to the
$B \to \phi K$ decay amplitudes.}
\label{table-3}}
\vspace{10mm}
\begin{center}
\begin{tabular}{|c||c|c||c|c||c|c||}
\hline Decay & \multicolumn{6}{c||}{$B^{\pm }\to \phi K^{\pm}$}
\\ \hline\hline ~~~Components~~~ &
\multicolumn{2}{c||}{twist-2~($10^{-4}$ GeV)~} &
\multicolumn{2}{c||}{higher-twist~($10^{-4}$ GeV)} &
\multicolumn{2}{c||}{ total~($10^{-4}$ GeV)} \\ \hline\hline
Amplitudes & ~~~~Re~~~~~ & ~~~Im~~~ & ~~~~~Re~~~~~~ & ~~~Im~~~ 
& ~~~~Re~~~~~ & ~~~Im~~~ \\ \hline\hline
$f_{\phi }F_{e}^{P}$ & 9.48 & --- & 27.71 & --- & 37.19 & --- \\
\hline $M_{e}^{P}$ & -3.30 & 2.44 & 1.60 & -1.11 & -1.70 & 1.33 \\
\hline $f_{B}F_{a}^{P}$ & 0.07 & -0.01 & -2.57 & -15.47 & -2.50
& -15.48 \\
\hline $M_{a}^{P}$ & 0.06 & 0.41 & 0.30 & -0.25 & 0.36
& 0.16 \\
\hline $f_{B}F_{a}$ & -1.21  & 0.47 & 41.00 & 13.52 &
39.79 & 13.99 \\
\hline $M_{a}$ & 0.88 & -12.74 & -7.05 & 2.61 &
-6.17 & -10.13 \\ \hline
\end{tabular}

\vspace{20mm}

\begin{tabular}{|c||c|c||c|c||c|c||}
\hline Decay & \multicolumn{6}{c||}{$B^{0}\to \phi K^{0}$}
\\ \hline\hline
~~~Components & \multicolumn{2}{c||}{twist-2~($10^{-4}$ GeV)} &
\multicolumn{2}{c||}{higher-twist~($10^{-4}$ GeV)} &
\multicolumn{2}{c||}{ total~($10^{-4}$ GeV)} \\ \hline\hline
Amplitudes & ~~~~Re~~~~~ & ~~~Im~~~ & ~~~~~Re~~~~~~ & ~~~Im~~~ 
& ~~~~Re~~~~~ & ~~~Im~~~ \\ \hline\hline
$f_{\phi}F_{e}^{P}$ & 9.48 & --- & 27.71 & --- & 37.19 & --- \\ \hline
$M_{e}^{P}$ & -3.30 & 2.44 & 1.60 & -1.11 & -1.70 & 1.33 \\ \hline
$f_{B}F_{a}^{P}$ & 0.07 & -0.01 & -2.68 & -15.76 & -2.61 &
-15.77 \\ \hline
$M_{a}^{P}$ & -0.03 & 0.77 & 0.48 & -0.28 & 0.45
& 0.49 \\ \hline
\end{tabular}
\end{center}
\end{table}

\newpage

\begin{table}[ht]
\vspace*{0.5cm}
\caption{{  Enhancement effects in the  $B^{\pm} \to \phi K^{\pm}$
decay amplitudes}
\label{table-4}}
\vspace{10mm}
\begin{center}
\begin{tabular}{|c||c|c||c|c||}
\hline ~~~Scales & \multicolumn{2}{c||}{$\mu =t$} &
\multicolumn{2}{c||}{$\mu =2.5$ GeV} \\
\hline\hline Amplitudes & Re~($10^{-4}$ GeV) & Im~($10^{-4}$ GeV)
& Re~($10^{-4}$ GeV) & Im~($10^{-4}$ GeV) \\ \hline\hline
$f_{\phi}F_{e}^{P}$ & 37.19 & --- & 23.14 & --- \\ \hline
$M_{e}^{P}$& -1.70 & 1.33 & -1.05 & 0.62 \\ \hline
$f_{B}F_{a}^{P}$& -2.50 & -15.48 & 1.92 & -12.83 \\ \hline
$M_{a}^{P}$ & 0.36 & 0.16 & -0.05 & 0.19 \\ \hline
$f_{B}F_{a}$ & 39.79 & 13.99 & 37.73 & 13.14 \\ \hline
$M_{a}$ & -6.17 & -10.13 & -0.56 & -9.05 \\ \hline\hline 
Branching Ratio &
\multicolumn{2}{c||}{$9.8\times 10^{-6}$} &
\multicolumn{2}{c||}{$3.8\times 10^{-6}$} \\ 
(without Ann.) & \multicolumn{2}{c||}{$$} &
\multicolumn{2}{c||}{$$} \\ \hline
Branching Ratio &
\multicolumn{2}{c||}{$10.2 \times 10^{-6}$} &
\multicolumn{2}{c||}{$5.6\times 10^{-6}$} \\ 
(with Ann.) &  \multicolumn{2}{c||}{$$} &
\multicolumn{2}{c||}{$$} \\   \hline
\end{tabular}
\end{center}
\end{table}

\end{document}